\documentclass[12pt]{iopart}

\usepackage{iopams}
\usepackage{graphicx}
\usepackage{bm}
\usepackage{color}
\usepackage{cite}
\usepackage{hyperref}

\hypersetup{
    colorlinks=true,
    linkcolor=blue,
    citecolor=blue,
    urlcolor=blue
}

\newcommand{\dd}{\mathrm{d}}

\newcommand{\ii}{\mathrm{i}}

\begin{document}

\title[The Trachenko--Zaccone equation]{The Trachenko--Zaccone equation: nonlinear relaxation from glasses to complex systems}

\author{
Alessio Zaccone$^{1,2}$,
Valeriy V. Ginzburg$^{3}$,
Oleg V. Gendelman$^{4}$,
Sofia F. Mauro$^{5}$,
John C. Mauro$^{5}$
}

\address{$^1$ Department of Physics ``A. Pontremoli'', University of Milan, Milan, Italy}

\address{$^2$ Institute for Theoretical Physics, University of Göttingen, Göttingen, Germany}

\address{$^3$ Department of Chemical Engineering and Materials Science,
Michigan State University, East Lansing, MI, USA}

\address{$^4$ Faculty of Mechanical Engineering,
Technion – Israel Institute of Technology, Haifa, Israel}

\address{$^5$ Department of Materials Science and Engineering,
Materials Research Institute,
The Pennsylvania State University,
University Park, PA 16802, USA}

\ead{alessio.zaccone@unimi.it; ginzbur7@msu.edu; ovgend@technion.ac.il; sfm6115@psu.edu; jcm426@psu.edu}

\begin{abstract}
The Trachenko--Zaccone equation provides a compact nonlinear dynamical framework for describing non-Debye relaxation in disordered condensed matter. Originally developed to rationalize stretched- and compressed-exponential relaxation in liquids and glasses from the dynamics of interacting local relaxation events, the same equation has subsequently appeared in broader contexts, including polymer relaxation and nonlinear models of global population dynamics. This review retraces the conceptual development of the equation, with particular emphasis on its physical origin in Kostya Trachenko's feed-forward interaction mechanism, its mathematical structure, and its possible generalizations. We also include personal recollections of the work with Kostya Trachenko at Queen Mary University of London in August 2019, during which the equation emerged in essentially its present form. After reviewing applications to stress relaxation, glassy materials, polymer relaxation and population dynamics, we discuss future directions including time-dependent feedback parameters, coupled order parameters, heterogeneous and spatially resolved formulations, flux terms, network versions and stochastic extensions. The central theme of the review is that the Trachenko--Zaccone equation should be viewed not only as a model of glassy relaxation, but as a general nonlinear feedback equation with potential applications across complex systems.
\end{abstract}

\noindent{\it Dedicated to the memory of Prof. Kostya Trachenko.}

\vspace{2pc}

\maketitle


\section{Introduction}
\label{sec:introduction}

One of the enduring goals of theoretical physics is to identify simple mathematical equations capable of describing complex phenomena across widely different physical systems. History offers many examples in which equations originally introduced to explain a specific physical problem were later recognized as possessing a much broader range of applicability. The diffusion equation, the Langevin equation, the Ginzburg--Landau equation and the Fisher--Kolmogorov equation all transcended the systems that originally motivated their derivation to become paradigmatic models of nonequilibrium dynamics.

Relaxation phenomena constitute one of the most ubiquitous manifestations of nonequilibrium dynamics. They occur in systems ranging from structural glasses and supercooled liquids to polymers, granular materials, colloidal suspensions and metallic glasses, as well as in biological tissues, ecological populations and social systems. Despite the enormous diversity of microscopic mechanisms involved, the macroscopic evolution of these systems is often governed by remarkably simple dynamical laws.

Among the oldest and most persistent challenges in condensed matter physics is understanding why relaxation in disordered systems so often deviates from the simple exponential behaviour predicted by Debye theory \cite{Debye1929}. Instead, experiments reveal stretched-exponential, compressed-exponential and logarithmic relaxation over many decades in time. The stretched exponential was first introduced empirically by Kohlrausch in 1854 while studying the discharge of a Leyden jar \cite{Kohlrausch1854}, and was later popularized in the context of dielectric relaxation by Williams and Watts \cite{WilliamsWatts1970}. Despite its extraordinary success in describing experimental observations across a remarkable diversity of materials, its microscopic origin remained one of the longstanding open problems in condensed matter physics, as emphasized in the influential review by Phillips \cite{Phillips1996}.

A major conceptual advance came from the realization that the slowing down of relaxation can emerge dynamically from interactions between local relaxation events, rather than being imposed phenomenologically through a distribution of relaxation times. This physical picture, originally developed for supercooled liquids and glasses, eventually led to the formulation of a nonlinear evolution equation capable of reproducing the principal relaxation regimes observed experimentally from a single dynamical framework. As will be discussed throughout this review, the same mathematical structure has subsequently emerged in apparently unrelated contexts, suggesting that its significance extends well beyond its original condensed-matter motivation.

Indeed, one of the most intriguing developments over the last few years has been the progressive expansion of this framework into areas well outside glass physics. Independent theoretical developments have shown that essentially the same nonlinear dynamics governs structural relaxation in polymer glasses \cite{GinzburgGendelmanZaccone2024}, while more recent work has demonstrated that it can also provide an effective description of global population dynamics over the last twelve millennia \cite{ZacconeTrachenko2026}. Although the microscopic interpretation is entirely different in these systems, the underlying nonlinear mathematical structure remains unchanged. This remarkable convergence suggests that the equation introduced later in this review should be regarded not merely as a model of glass relaxation, but as a more general nonlinear evolution equation for cooperative dynamics in complex systems.

\begin{figure*}[t]
\centering
\includegraphics[width=0.98\textwidth]{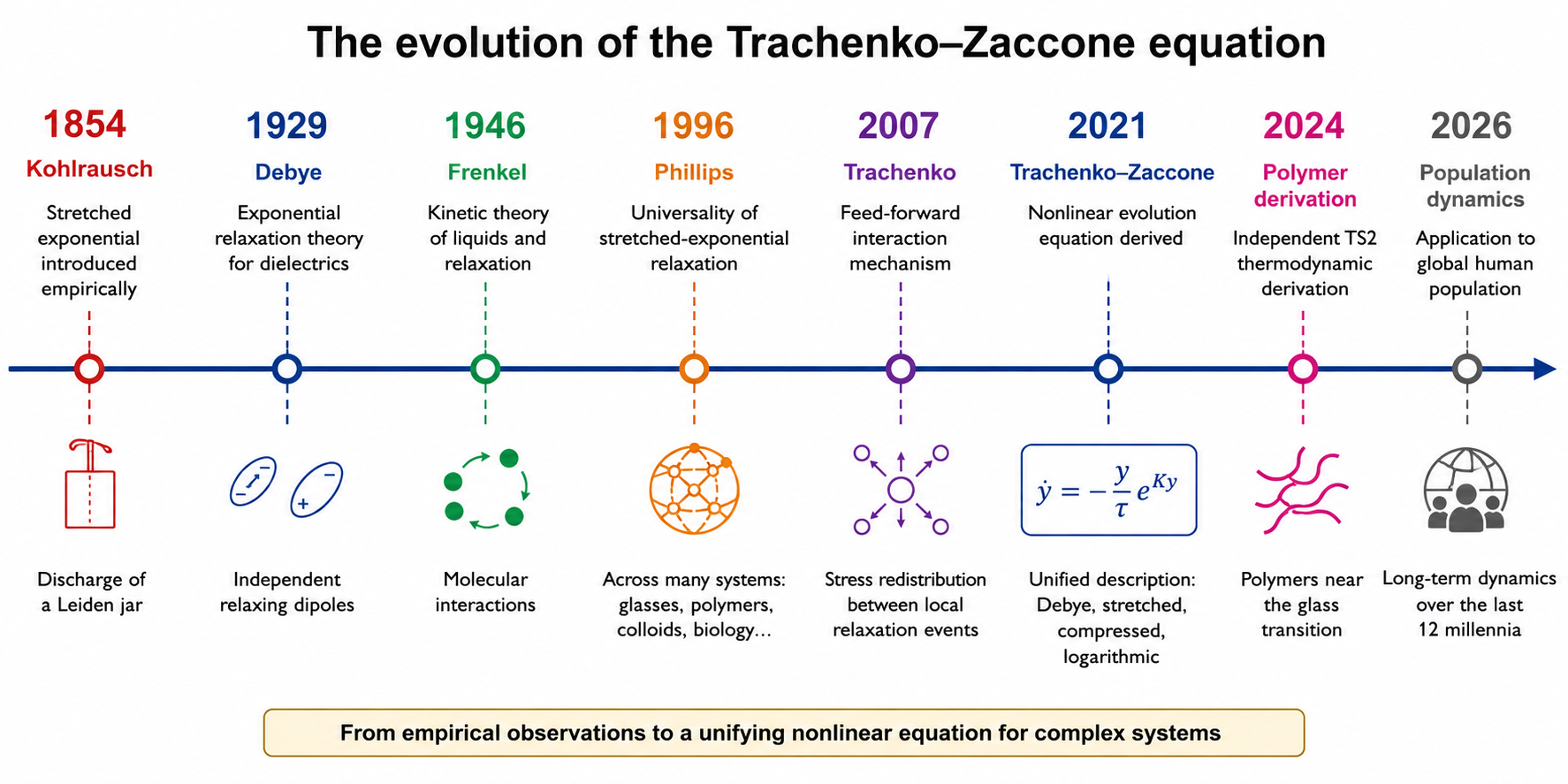}
\caption{
Historical evolution of the ideas leading to the Trachenko--Zaccone equation. Beginning with Kohlrausch's empirical stretched-exponential relaxation (1854) and Debye's theory of exponential relaxation (1929), progressively deeper physical understanding emerged through Frenkel's kinetic theory of liquids \cite{Frenkel1946}, Phillips' demonstration of the universality of stretched-exponential relaxation, and Trachenko's feed-forward interaction mechanism for interacting local relaxation events. These developments culminated in the formulation of the Trachenko--Zaccone equation in 2021, which unified Debye, stretched-exponential, compressed-exponential and logarithmic relaxation within a single nonlinear dynamical framework. Subsequent work demonstrated an independent thermodynamic derivation for polymer glasses and, more recently, the applicability of the same mathematical structure to nonlinear population dynamics, illustrating its potential as a general nonlinear evolution equation for complex systems.
}
\label{fig:timeline}
\end{figure*}

The historical evolution leading to these developments is summarized in Fig.~\ref{fig:timeline}.

The aim of this article is not simply to review the existing literature surrounding the Trachenko--Zaccone equation, but to retrace its conceptual development from its origins in the physics of structural relaxation to its emerging role as a general framework for nonlinear dynamics. We discuss the physical ideas that motivated its derivation, the mathematical properties of the resulting nonlinear equation, its applications to glasses, polymers and population dynamics, and several possible future extensions including heterogeneous media, reaction--diffusion formulations, interacting populations, stochastic dynamics and coupled nonlinear fields.

This article is dedicated to the memory of Konstantin (Kostya) Trachenko, whose originality of thinking, physical intuition and enthusiasm inspired much of the work described here. Some sections include personal recollections of the discussions that took place during Alessio Zaccone's visit to Queen Mary University of London in August 2019, when the nonlinear equation first emerged from attempts to formulate the feed-forward interaction mechanism in its simplest mathematical form. Looking back only a few years later, it is remarkable how an equation originally conceived to understand slow relaxation in glasses has progressively evolved into a framework with potential applications across disciplines. This evolution reflects Kostya's distinctive scientific style: searching for simple physical principles whose mathematical formulation naturally reveals unexpected universality.

\section{Historical background}
\label{sec:historical}

The modern understanding of structural relaxation has developed over more than one and a half centuries, progressing from empirical observations to increasingly microscopic theoretical descriptions. Although many phenomenological models were proposed during this period, a unified physical explanation of non-Debye relaxation remained elusive for decades. This historical evolution ultimately culminated in the feed-forward interaction mechanism proposed by Trachenko, which provided the physical basis for the nonlinear evolution equation discussed in the following sections. In this section, we briefly review the principal milestones that led to this development.

\subsection{Debye relaxation}

The modern theoretical description of relaxation originates with Debye's theory of dielectric polarization \cite{Debye1929}. In its simplest form, the relaxation of an observable $y(t)$ toward equilibrium is governed by the linear differential equation

\begin{equation}
\frac{\dd y}{\dd t}
=
-\frac{y}{\tau},
\label{eq:Debye}
\end{equation}

whose solution is the familiar exponential law

\begin{equation}
y(t)=y_0 e^{-t/\tau}.
\end{equation}

The Debye equation represents one of the simplest examples of linear irreversible dynamics. Its central assumption is that the relaxation rate is constant throughout the evolution and independent of the instantaneous state of the system.

Although extraordinarily successful for dilute systems whose relaxing units
evolve independently, the Debye model neglects cooperative interactions
between local relaxing regions. As interactions become increasingly important,
particularly in dense liquids, supercooled liquids and glasses, the assumption
of a constant relaxation rate becomes inadequate. Experiments consistently
reveal significant deviations from exponential behaviour, indicating that the
relaxation rate itself evolves during the relaxation process.

\subsection{The Kohlrausch stretched exponential}

Long before Debye's work, Kohlrausch had already observed that the discharge of a Leyden jar could be described much more accurately by the empirical expression

\begin{equation}
y(t)
=
y_0
\exp
\left[
-
\left(
\frac{t}{\tau}
\right)^\beta
\right],
\qquad
0<\beta<1,
\label{eq:KWW}
\end{equation}

now universally known as the stretched-exponential or Kohlrausch--Williams--Watts (KWW) law \cite{Kohlrausch1854,WilliamsWatts1970}.

 Remarkably, this deceptively simple expression has proved extraordinarily successful in describing relaxation across an enormous variety of systems, including structural glasses, polymers, colloids, spin glasses, granular materials and biological systems. Yet, despite more than one and a half centuries of investigation, no generally accepted microscopic derivation emerged. Understanding why such a universal empirical law appears in so many apparently unrelated systems became one of the central challenges of glass physics.

\begin{quote}
\textbf{Historical perspective.}
It is remarkable that an empirical function introduced in 1854 remained without a generally accepted microscopic derivation for more than 160 years. During this period, numerous theoretical approaches were proposed, including distributions of activation energies, distributions of relaxation times, trap models, heterogeneous dynamics, continuous-time random walks, fractal models and facilitated dynamics. While each successfully described particular aspects of experimental observations, none provided a broadly accepted microscopic origin for stretched-exponential relaxation. The search for a microscopic origin of stretched-exponential relaxation became one of the classical problems of glass physics.
\end{quote}

\subsection{Universality of stretched-exponential relaxation}

The remarkable universality of stretched-exponential relaxation was extensively reviewed by Phillips \cite{Phillips1996}, who collected evidence from hundreds of experimental systems spanning condensed matter physics, chemistry and biology. This review established beyond doubt that stretched-exponential relaxation is not a peculiarity of glasses but a generic feature of many complex systems.

The extraordinary breadth of these observations strongly suggested that the origin of stretched relaxation should not be sought in material-specific microscopic details, but rather in generic collective mechanisms capable of producing nonlinear slowing down.

Phillips argued that the extraordinary universality of stretched-exponential relaxation strongly suggests the existence of generic physical principles largely independent of the microscopic details of individual materials. This perspective considerably shifted the emphasis of the field: rather than searching for material-specific explanations, attention increasingly focused on identifying universal mechanisms capable of producing cooperative nonlinear relaxation.

\subsection{Feed-forward interaction mechanism}

An important advance came with Trachenko's feed-forward interaction mechanism \cite{Trachenko2007}. The central idea is that local relaxation events occurring within a stressed liquid are not independent but interact elastically over increasing distances as the relaxation proceeds.

Indeed, the key conceptual step introduced by Trachenko was to abandon the traditional picture in which local relaxation events occur independently. Instead, each local relaxation event was recognized to modify the surrounding elastic stress field, thereby changing the activation barriers encountered by subsequent relaxation events. Relaxation therefore becomes intrinsically self-modifying: the dynamics alters the very landscape in which future dynamics takes place. As relaxation proceeds, the range over which local relaxation events interact grows because the liquid progressively acquires elastic coherence over increasing distances. Consequently, later relaxation events experience larger activation barriers than earlier ones, naturally producing a progressive slowing down of the overall relaxation dynamics without invoking any ad hoc distribution of relaxation times.
This feed-forward interaction mechanism represented a profound conceptual departure from previous approaches. Rather than assuming heterogeneity from the outset, heterogeneous relaxation emerges dynamically from the cooperative interactions between local relaxation events themselves.

Each local relaxation event redistributes stress onto neighbouring regions, thereby modifying the activation barriers of subsequent relaxation events. Consequently, the relaxation rate is no longer constant but evolves
dynamically during relaxation through a feed-forward interaction mechanism.

This physical picture represented a significant departure from earlier phenomenological approaches. Instead of assuming a distribution of relaxation times, the slowing down emerges dynamically from the interaction between local relaxation events themselves.

As will be discussed in the following sections, reformulating this microscopic mechanism into a nonlinear differential equation ultimately led to what is now known as the Trachenko--Zaccone equation.

\subsection{The stretched exponential for non-Debye relaxation}
\label{subsec:stretched_exponential_mauro}

As introduced in section~2.2, most real relaxation processes do not
display perfect Debye behavior and are commonly described by the
stretched-exponential or Kohlrausch--Williams--Watts (KWW) function,
equation~(\ref{eq:KWW}). The stretched exponential is an empirically proposed
function introduced by Kohlrausch in 1854 to describe the charge
relaxation in a Leyden jar \cite{Kohlrausch1854}
and was popularized by Williams and Watts in their 1970 paper
\cite{WilliamsWatts1970}. Independent of chemical composition, it is
universally capable of describing homogeneous glass relaxation
\cite{Phillips1996,MauroMauro2018}. The stretched exponential is also
commonly used to fit stress and structural relaxation of polymers and
other viscoelastic materials. Other applications include dielectric
response and molecular luminescence.

The KWW function contains two free parameters: the stretching exponent
$\beta$ and the relaxation time $\tau$. The dimensionless stretching
exponent is constrained to values $0<\beta\leq1$ for
stretched-exponential relaxation, where values close to unity approach
simple exponential decay. Less exponential processes have smaller
values of $\beta$.

Multiple physical models have successfully attempted to reproduce the empirical stretched
exponential function. The first and most well-known of these, the ``diffusion in traps'' model,
was developed by Grassberger and Procaccia in 1982 \cite{GrassbergerProcaccia1982} and extended by Phillips in 1994
\cite{Phillips1994}. In this experiment, the stretched exponential emerges from the decay of particle density following
diffusion around a random distribution of stationary traps. Regions having fewer traps lead to longer
particle decay times, giving rise to the stretched exponential functional form.

The model by Phillips extends these findings, showing that for microscopically
homogeneous systems, the stretching exponent is related to the dimensionality of relaxation
pathways, $d^{*}$ \cite{Phillips1994,MacdonaldPhillips2005,Phillips2006}:

\begin{equation}
\beta
=
\frac{d^{*}}{d^{*}+2}.
\label{eq:mauro_phillips}
\end{equation}

In turn, dimensionality of relaxation pathways $d^{*}$ is calculated as $d^{*}=fd$, where $f$
is the fraction of pathways that are activated in the system, and $d$ is the dimensionality of
the system. It follows that structural relaxation of a three-dimensional, fully activated
system, with $d=3$ and $f=1$, leads to a stretching exponent value $\beta=3/5$. To maximize the
rate of entropy production in the relaxation process, long-range and short-range
contributions must be split evenly. When only long-range relaxation processes occur in a
three-dimensional system ($d=3$), the value of $f$ is $1/2$, leading to a fractal dimensionality
of $d^{*}=3/2$ and a stretching exponent of $\beta=3/7$. $\beta$ has been shown to bifurcate into the
values of $3/5$ and $3/7$ for stress and structural relaxation, respectively. In other words,
both short-range and long-range relaxation pathways are active under an applied load
(stress relaxation), whereas only long-range contributions occur under no applied load
(structural relaxation) \cite{NaumisPhillips2012,PotuzakWelchMauro2011}. Another common value of
$\beta$ is $1/2$, which is obtained for two-dimensional ($d=2$), fully activated ($f=1$) materials.
The stretching exponent is therefore linked to physically meaningful parameters, as a smaller
value of $\beta$ corresponds to lower dimensionality of relaxation pathways
\cite{Phillips1996}.

A different route to broad relaxation spectra has recently been proposed
within the ``switchback'' model developed by Medvedev and co-workers
\cite{Medvedev2023Switchback,MedvedevYungbluthSavoieCaruthers2024}.
In this picture, the overall relaxation results from multiple coupled
processes operating over multiple relaxation times. Their cumulative response
can generate broad KWW-like relaxation, while an appropriate limiting case
recovers single-exponential Debye-like behaviour. This provides another
example in which a non-Debye relaxation law can emerge from the collective
superposition of elementary dynamical processes rather than from an assumed
KWW functional form.

A recently proposed physical origin of the stretched exponential function is entropy
relaxation in a system of microstates represented by an energy landscape having $N$ number of
locally stable inherent structures, located within their own basins
\cite{MauroSmithMauro2026}. This system represents the
relaxation of a kinetically constrained system, such as a glass, towards equilibrium. As the glass
relaxes, the configurational degrees of freedom that were frozen during vitrification are gradually
restored. While derived for glass relaxation processes, this energy landscape construction may be
used to describe any Markovian system where each transition is independent of the previous.

The rates of change of occupation probability $p_i$ of basin $i$ form a series of master equations
of the form:

\begin{equation}
\frac{\mathrm{d}p_i}{\mathrm{d}t}
=
(\mathrm{Flow\ in})-(\mathrm{Flow\ out})
=
k_{\rm tr}\left(
\sum_{j\neq i}p_j
-
\sum_{j\neq i}p_i
\right),
\label{eq:mauro_master}
\end{equation}

where

\begin{equation}
k_{\rm tr}
=
\nu
\exp\left(
-\frac{U^{*}}{k_{\rm B}T}
\right),
\label{eq:mauro_transition_rate}
\end{equation}

in which $\nu$ is the vibrational frequency (i.e., the frequency of attempts to exit the basin) and
the exponential term is the Boltzmann probability factor (i.e., the probability of successful activation of
the system over the energy barrier $U^{*}$ for inter-basin transitions). $k_{\rm tr}$ is assumed to be constant,
capturing the transition from non-ergodic to ergodic states in glass relaxation, for which there is no
associated latent heat. $U^{*}$ is also held constant to isolate the impact of entropy.

Such master equations describing the evolution of occupation probabilities of each basin are
constructed for two limiting cases: a completely connected topology where the system may relax
into any basin regardless of proximity and a sequential topology where the system must move among
adjacent basins. Once these are solved analytically, the Gibbs entropy function,

\begin{equation}
S(t)
=
-k_{\rm B}
\sum_{i=1}^{N}
p_i\ln p_i,
\label{eq:mauro_gibbs_entropy}
\end{equation}

relates probability to entropy production. Finally, the relaxation function is written as normalized
entropy:

\begin{equation}
g(t)
=
\frac{S(t)-S_{\infty}}
{S_0-S_{\infty}}
=
1-\frac{S(t)}{k_{\rm B}\ln N},
\qquad
0\leq g(t)\leq 1.
\label{eq:mauro_normalized_entropy}
\end{equation}

which approximates the stretched exponential for both limiting cases. For a sequential topology, $\tau$
increases monotonically with number of basins $N$, while $\beta$ decays monotonically with $N$, becoming
increasingly nonexponential. For a fully connected topology, the trends are opposite: $\tau$ decays and
$\beta$ grows with increasing $N$. These differences illustrate the nature of entropy relaxation as
dependent primarily on available transition points when all else remains the same
\cite{MauroSmithMauro2026}.

While the stretched exponential fits well to both limiting cases, the fully connected energy
landscape provides a better reproduction of the function. This calls attention to a discrepancy
between the stretched exponential function and the analytical solution derived for the sequential,
one-dimensional topology. In this energy landscape, probability flows sequentially, not
simultaneously, into each basin. In contrast, the stretched exponential function considers each
relaxation mode to occur simultaneously, which is one reason why the Prony series approximation
discussed below can reproduce the stretched exponential: the series is composed of simple
exponential decay functions starting at $t=0$ \cite{MauroMauro2018}. While each transition occurs on its own time scale,
every transition begins at once. For a sequential energy landscape, this discrepancy becomes more
significant as $N$ increases, resulting in greater error. The opposite is true for the fully connected
topology, which fits the stretched exponential better as the number of basins increases.

For ease of computation, a Prony series may be used to approximate the stretched
exponential with a discrete sum of simple exponentials,

\begin{equation}
\exp\left[
-\left(\frac{t}{\tau}\right)^{\beta}
\right]
\approx
\sum_{i=1}^{N}
w_i
\exp\left[
-k_i\left(\frac{t}{\tau}\right)
\right],
\label{eq:mauro_prony}
\end{equation}

with weighting factors $w_i$ that sum to 1. Values of the coefficients $w_i$ and $k_i$ have been optimized by
Mauro and Mauro \cite{MauroMauro2018} for the three critical values of $\beta$ and increasing
numbers of terms in the Prony series ($N$). Terms with higher values of $k_i$ approximate
behavior at short times, while terms with lower values of $k_i$ model the ``fat tail'' of the stretched
exponential, or the area beneath the tail at long times. The quality of fitting increases exponentially
with increasing $N$ and decreases linearly with decreasing $\beta$ for $\beta>0.3$
\cite{MauroMauro2018}.

The Prony series is less capable of approximating the stretched exponential at short times,
when the stretched exponential decays much faster than a simple exponential can model. Another
inconsistency is that the slope of the Prony series at $t=0$ is finite and can never reproduce the
divergent, infinite slope of the stretched exponential at $t=0$. However, for the purposes of reducing
the computational cost of modeling nonexponential non-Debye relaxation, the Prony series is a
useful tool.

\section{Origins of the Trachenko--Zaccone equation}
\label{sec:origins}

The nonlinear evolution equation reviewed in this article did not originate as an attempt to devise another phenomenological fitting function for non-Debye relaxation. Rather, it emerged from a more specific physical question: can the feed-forward interaction mechanism introduced by Trachenko for local relaxation events in viscous liquids be written as a closed dynamical equation for the macroscopic relaxation variable?

This question became the focus of discussions between Alessio Zaccone and Kostya Trachenko during Zaccone's visit to Queen Mary University of London in August 2019. At that time, two apparently separate observations provided the main motivation. On the theoretical side, the feed-forward mechanism developed by Trachenko had shown how elastic interactions between successive local relaxation events could generate progressively increasing activation barriers and hence slow non-exponential dynamics \cite{Trachenko2007}. 

\begin{figure}[ht]
\centering
\includegraphics[width=0.88\textwidth]{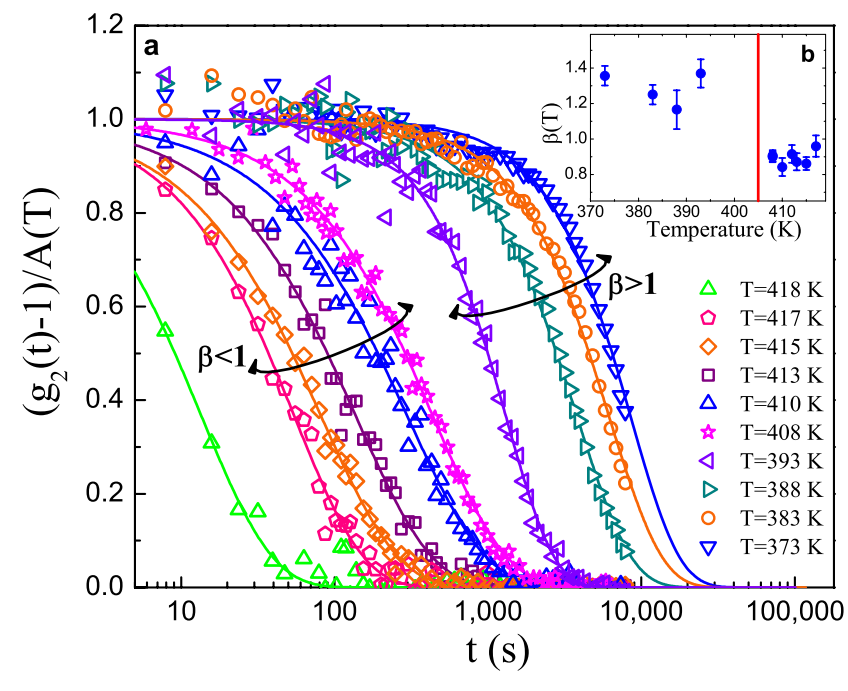}
\caption{Experimental motivation for a unified description of stretched- and
compressed-exponential relaxation. Normalized intensity-correlation functions
measured by X-ray photon correlation spectroscopy in a metallic glass at
different temperatures. The relaxation changes from stretched-exponential
behaviour ($\beta<1$) at higher temperatures to compressed-exponential
behaviour ($\beta>1$) at lower temperatures. The inset shows the corresponding
temperature dependence of the exponent $\beta$, with the vertical line
indicating the glass-transition region. The continuous evolution of $\beta$
across unity provides an important motivation for seeking a single nonlinear
dynamical equation capable of describing both relaxation regimes.
Reproduced from Ruta \textit{et al.} \cite{Ruta2012} with permission of the American Physical Society.}
\label{fig:ruta_SER_CER}
\end{figure}

On the experimental side, X-ray photon-correlation measurements on metallic
glasses had revealed something particularly intriguing: the relaxation
changes from stretched-exponential behaviour, with $\beta<1$, to
compressed-exponential behaviour, with $\beta>1$, upon entering the glassy
state \cite{Ruta2012}. As illustrated in figure~\ref{fig:ruta_SER_CER}, the
experiment does not merely reveal two isolated relaxation laws, but a
temperature-dependent evolution of the effective exponent across $\beta=1$.
This observation raises a natural question: can stretched and compressed
relaxation emerge as two regimes of the same underlying dynamical equation?

These observations presented a challenge to existing descriptions. Broad distributions of relaxation times can naturally generate stretched exponentials, and diffusion-to-traps models can yield particular values of the stretching exponent, such as the well-known $\beta=3/5$ result in three dimensions \cite{Phillips1996}. Mode-coupling approaches provide a microscopic framework for the slow $\alpha$ relaxation of supercooled liquids, while elastic-dipole and micro-collapse models can produce compressed exponentials with particular exponents. The experimental situation, however, pointed to a continuous range of values of $\beta$ on both sides of unity rather than to a small set of distinguished values. This suggested that stretched and compressed relaxation might be two regimes of a more general dynamical law rather than fundamentally unrelated phenomena.

The discussions in 2019 therefore concentrated on identifying the minimal rate equation implied by the feed-forward picture. The key step was to recognize that the activation barrier itself evolves as local relaxation events accumulate. Since the elementary event probability is Arrhenius, a barrier that depends linearly on the relaxation state immediately generates an exponential nonlinearity in the rate equation. This observation ultimately led to the equation now referred to as the Trachenko--Zaccone (TZ) equation \cite{TrachenkoZaccone2021}.

\subsection{Local relaxation events and elastic feed-forward interactions}
\label{subsec:LRE}

Following the physical picture introduced by Orowan \cite{Orowan1952} and Goldstein \cite{Goldstein1969} and developed for viscous liquids by Trachenko, structural relaxation proceeds through localized rearrangements, which we refer to as local relaxation events (LREs). Under an applied external stress, the statistically favoured events are \emph{concordant}: their local strain is aligned so as to reduce the applied stress.

An important consequence follows from mechanical equilibrium. Once a concordant event has taken place, the relaxed region supports less of the externally imposed stress. The stress released locally must therefore be redistributed to other regions that have not yet relaxed. A region undergoing a later LRE consequently experiences an additional stress $\Delta p$ generated by earlier events.

The activation barrier $V$ of an LRE is related to the mechanical work required to achieve the local rearrangement. In a cage picture, for example, an atom or molecule must push aside its neighbours in order to escape its local environment. If $q$ denotes the local cage volume, an additional stress $\Delta p$ changes the activation barrier by

\begin{equation}
\Delta V=\int \Delta p \, d q .
\label{eq:deltaV_general}
\end{equation}

Introducing a characteristic activation or cage volume $q_0$, this becomes, to leading order,

\begin{equation}
\Delta V\simeq q_0\Delta p ,
\label{eq:deltaV}
\end{equation}

and hence

\begin{equation}
V(n)=V_0+q_0\Delta p(n),
\label{eq:V_Deltap}
\end{equation}

where $V_0$ is the activation barrier before the feed-forward contribution is included and $n$ is the number of LREs that have already occurred.

The dependence of $\Delta p$ on $n$ can be obtained explicitly from elasticity. Consider a prospective LRE at the centre of a sphere of radius $d_{\rm el}$. The quantity

\begin{equation}
d_{\rm el}=c\tau_{\alpha}
\label{eq:elasticity_length}
\end{equation}

is the elasticity length, where $c$ is the relevant sound velocity and $\tau_{\alpha}$ is the structural relaxation time. It gives the distance over which elastic stress can propagate before being interrupted by another relaxation event.

Let $\Delta p_i(r)$ be the additional stress at the central region generated by a previous concordant LRE located at distance $r$. In three-dimensional elasticity, the far-field stress produced by a localized rearrangement decays as

\begin{equation}
\Delta p_i(r)\propto \frac{1}{r^3}.
\end{equation}

Taking $\Delta p_i(d_0/2)=\Delta p_0$, where $d_0$ is the characteristic size of a local rearranging region, gives

\begin{equation}
\Delta p_i(r)=
\Delta p_0
\left(\frac{d_0}{2r}\right)^3 .
\label{eq:stress_decay}
\end{equation}

If the instantaneous density of completed LREs is $\rho$, the total additional stress acting on the central region is obtained by summing the elastic contribution of all previous LREs contained within the elasticity sphere:

\begin{equation}
\Delta p
=\rho
\int_{d_0/2}^{d_{\rm el}}
4\pi r^2
\Delta p_i(r)\,d r .
\label{eq:stress_integral}
\end{equation}

Substitution of equation~(\ref{eq:stress_decay}) gives

\begin{eqnarray}
\Delta p
&=&
4\pi\rho\Delta p_0
\frac{d_0^3}{8}
\int_{d_0/2}^{d_{\rm el}}
\frac{d r}{r}
\nonumber \\
&=&
\frac{\pi}{2}
\rho\Delta p_0 d_0^3
\ln\left(\frac{2d_{\rm el}}{d_0}\right).
\label{eq:stress_result}
\end{eqnarray}

We now introduce $n(t)$, the number of LREs that have occurred inside the elasticity sphere at time $t$, and $n_r$, the total number that will have occurred once the imposed perturbation has completely relaxed. Their corresponding densities are

\begin{equation}
\rho(t)=\frac{6n(t)}{\pi d_0^3},
\qquad
\rho_r=\frac{6n_r}{\pi d_0^3}.
\label{eq:LRE_density}
\end{equation}

Therefore,

\begin{equation}
\rho(t)=\rho_r\frac{n(t)}{n_r}.
\end{equation}

Using equations~(\ref{eq:V_Deltap}) and (\ref{eq:stress_result}), the activation barrier assumes the particularly simple form

\begin{equation}
V(n)=
V_0
+
V_1\frac{n}{n_r},
\label{eq:Vn}
\end{equation}

where

\begin{equation}
V_1
=\frac{\pi}{2}
\rho_rq_0\Delta p_0d_0^3
\ln\left(\frac{2d_{\rm el}}{d_0}\right).
\label{eq:V1}
\end{equation}

Equation~(\ref{eq:Vn}) is the microscopic origin of the nonlinearity that appears below. Earlier relaxation events progressively increase the activation barrier encountered by later events. The process therefore possesses an intrinsic form of memory: the instantaneous rate depends on how far the relaxation has already progressed.

It is worth stressing that this memory does not need to be introduced through an explicit memory kernel. It is encoded in the evolving state variable $n(t)$ itself.

\subsection{Population balance for local relaxation events}
\label{subsec:population_balance}

The next step follows from a simple population-balance argument. The instantaneous number of new LREs per unit time must be proportional to two factors. The first is the number of regions that have not yet relaxed,

\begin{equation}
n_r-n(t).
\end{equation}

The second is the thermally activated probability for an LRE to occur,

\begin{equation}
P_{\rm LRE}=
\exp\left[-\frac{V(n)}{k_{\rm B}T}\right].
\label{eq:event_probability}
\end{equation}

Introducing a microscopic attempt time $\tau_0$, the rate equation can consequently be written as

\begin{equation}
\frac{d n}{d t}=
\frac{1}{\tau_0}
\left[n_r-n(t)\right]
\exp\left[-\frac{V(n)}{k_{\rm B}T}\right].
\label{eq:dn_dt}
\end{equation}

It is useful to introduce the fraction of LREs that have already occurred,

\begin{equation}
q(t)=\frac{n(t)}{n_r},
\qquad
0\leq q\leq 1.
\label{eq:q_definition}
\end{equation}

Combining equations~(\ref{eq:Vn}) and (\ref{eq:dn_dt}) gives

\begin{equation}
\frac{d q}{d t}=
\frac{1}{\tau_0}
\exp\left(-\frac{V_0}{k_{\rm B}T}\right)
(1-q)
\exp(-Kq),
\label{eq:q_SER}
\end{equation}

where the dimensionless quantity

\begin{equation}
K=\frac{V_1}{k_{\rm B}T}
\label{eq:Kmicro}
\end{equation}

measures the strength of the elastic feed-forward interaction relative to thermal energy.

Equation~(\ref{eq:q_SER}) already contains the essential physics of stretched-exponential relaxation. As more LREs occur, $q$ increases, the activation barrier increases, and the factor $\exp(-Kq)$ progressively suppresses the rate of subsequent events. The dynamics therefore slows itself down.

\subsection{From the LRE population balance to the TZ equation}
\label{subsec:TZ_derivation}

To express the dynamics in the conventional language of relaxation, it is more convenient to consider the fraction of the initial perturbation that remains unrelaxed,

\begin{equation}
y(t)=1-q(t).
\label{eq:y_definition}
\end{equation}

Thus $y(0)=1$ and $y(t)\rightarrow 0$ for $t\rightarrow\infty$. Since $q=1-y$, equation~(\ref{eq:q_SER}) becomes

\begin{eqnarray}
\frac{\dd y}{\dd t}
&=&
-\frac{1}{\tau_0}
\exp\left(-\frac{V_0}{k_{\rm B}T}\right)
y
\exp[-K(1-y)]
\nonumber \\
&=&
-\frac{1}{\tau_0}
\exp\left[-\frac{V_0+V_1}{k_{\rm B}T}\right]
y\exp(Ky).
\label{eq:TZ_before_tau}
\end{eqnarray}

Defining the effective relaxation time

\begin{equation}
\tau=
\tau_0
\exp\left(\frac{V_0+V_1}{k_{\rm B}T}\right),
\label{eq:tau_TZ}
\end{equation}

we obtain the Trachenko--Zaccone equation,

\begin{equation}
\frac{d y}{d t}
=-\frac{y}{\tau}\exp(Ky).
\label{eq:TZ}
\end{equation}

The exponential nonlinearity in equation~(\ref{eq:TZ}) is therefore not postulated phenomenologically. It follows from two elementary ingredients: an activation barrier modified by the accumulated elastic effect of previous relaxation events, and the Arrhenius dependence of the event probability on that activation barrier.

The equation contains two parameters with transparent meanings. The time $\tau$ sets the overall relaxation time scale, whereas $K$ is a dimensionless feedback or cooperativity parameter. For $K>0$, earlier relaxation events hinder subsequent ones, producing a progressively decreasing effective rate and hence stretched relaxation.

An equivalent way of seeing this is to divide equation~(\ref{eq:TZ}) by $y$:

\begin{equation}
-\frac{d \ln y}{d t}=
\frac{1}{\tau}\exp(Ky).
\label{eq:instantaneous_rate}
\end{equation}

The quantity on the left-hand side is the instantaneous logarithmic relaxation rate. In Debye theory it is simply the constant $1/\tau$. In the TZ equation it is instead a function of the instantaneous state $y(t)$. Non-Debye relaxation therefore arises because the system continuously modifies its own relaxation rate.

\subsection{Debye relaxation as the non-interacting limit}
\label{subsec:Debye_limit}

The microscopic origin of the Debye limit is particularly transparent. At sufficiently high temperature the structural relaxation time becomes short and the elasticity length

\begin{equation}
d_{\rm el}=c\tau_{\alpha}
\end{equation}

approaches the scale of a local rearranging region. Long-range elastic communication between LREs then disappears, the logarithmic factor in equation~(\ref{eq:V1}) becomes negligible, and

\begin{equation}
V_1\rightarrow 0,
\qquad
K\rightarrow 0.
\end{equation}

Equation~(\ref{eq:TZ}) consequently reduces to

\begin{equation}
\frac{d y}{d t}
=-\frac{y}{\tau},
\end{equation}

with solution

\begin{equation}
y(t)=\exp(-t/\tau).
\label{eq:Debye_from_TZ}
\end{equation}

Debye relaxation is thus not a separate law added to the model: it is the high-temperature, effectively non-interacting limit of the same microscopic equation.

\subsection{From feed-forward slowing down to avalanche acceleration}
\label{subsec:CER_derivation}

The experimental observation of compressed-exponential relaxation in metallic glasses required one further conceptual step. The Orowan--Goldstein feed-forward mechanism assumes that elastic quasi-equilibrium is established after successive relaxation events. Under these conditions, previous concordant events increase the load on regions that have not yet relaxed, leading to the increasing barrier of equation~(\ref{eq:Vn}).

Deep in the glassy state, however, relaxation can instead proceed through intermittent avalanche-like events. A local rearrangement occurring in a highly stressed non-equilibrium configuration may destabilize neighbouring regions and make subsequent rearrangements easier rather than harder. The sign of the feedback is then reversed. Microscopically, this can be represented as

\begin{equation}
V(n)
=V_0
-V_1^{\prime}\frac{n}{n_r},
\qquad
V_1^{\prime}>0.
\label{eq:V_CER}
\end{equation}

The corresponding population-balance equation becomes

\begin{equation}
\frac{d q}{d t}
=
\frac{1}{\tau_0}
\exp\left(-\frac{V_0}{k_{\rm B}T}\right)
(1-q)
\exp(K^{\prime}q),
\label{eq:q_CER}
\end{equation}

where

\begin{equation}
K^{\prime}
=
\frac{V_1^{\prime}}{k_{\rm B}T}>0.
\end{equation}

Transforming again to the unrelaxed fraction $y=1-q$ gives

\begin{equation}
\frac{d y}{d t}
=
-\frac{1}{\tau_0}
\exp\left[
-\frac{V_0-V_1^{\prime}}{k_{\rm B}T}
\right]
y\exp\left(-K^{\prime}y\right).
\label{eq:cer_y}
\end{equation}

Introducing

\begin{equation}
\tau
=
\tau_0
\exp\left(\frac{V_0-V_1^{\prime}}{k_{\rm B}T}\right)
\end{equation}

and identifying

\begin{equation}
K=-K^{\prime}<0,
\end{equation}

one recovers exactly the same TZ equation,

\begin{equation}
\frac{d y}{d t}
=
-\frac{y}{\tau}\exp(Ky).
\end{equation}

The sign of a single parameter therefore distinguishes two physically different types of cooperative dynamics. For $K>0$, previous events hinder subsequent relaxation and stretched-exponential behaviour is obtained. For $K=0$, the events become effectively independent and ordinary Debye relaxation is recovered. For $K<0$, previous events facilitate subsequent ones, producing accelerated, compressed-exponential relaxation.

This unification was one of the central motivations for introducing the TZ equation. The stretched and compressed regimes are not obtained by postulating two different fitting functions. They arise from the same nonlinear dynamical law, with the sign of the feedback parameter encoding whether previous events hinder or facilitate subsequent relaxation.

\subsection{Physical meaning of the nonlinear feedback}
\label{subsec:feedback_interpretation}

Equation~(\ref{eq:TZ}) can be viewed in a broader way. Defining a state-dependent effective relaxation rate

\begin{equation}
\Gamma(y)
=
\frac{1}{\tau}\exp(Ky),
\label{eq:Gamma}
\end{equation}

the TZ equation assumes the deceptively simple form

\begin{equation}
\frac{d y}{d t}
=
-\Gamma(y)y.
\end{equation}

The difference from ordinary Debye relaxation lies entirely in the fact that $\Gamma$ is no longer a constant. The relaxation variable changes its own rate of evolution through nonlinear feedback.

This interpretation also makes clear why the mathematical structure can survive far beyond the particular microscopic derivation given above. The glass-physics derivation identifies $K$ with elastic cooperativity between LREs, but equation~(\ref{eq:TZ}) only requires a rate that depends exponentially on a state-dependent barrier or effective propensity. In other physical contexts, the microscopic origin of $K$ may be entirely different while the nonlinear mathematical structure remains the same.

This observation, which was not apparent when the equation was first formulated, ultimately opened the way to its later derivation within polymer statistical thermodynamics and, more recently, to its application to global population dynamics. In this sense, the LRE derivation provides not only a microscopic theory of non-Debye relaxation, but also the first physical realization of a more general nonlinear rate-feedback principle.

\begin{figure}[ht]
\centering
\includegraphics[width=0.98\textwidth]
{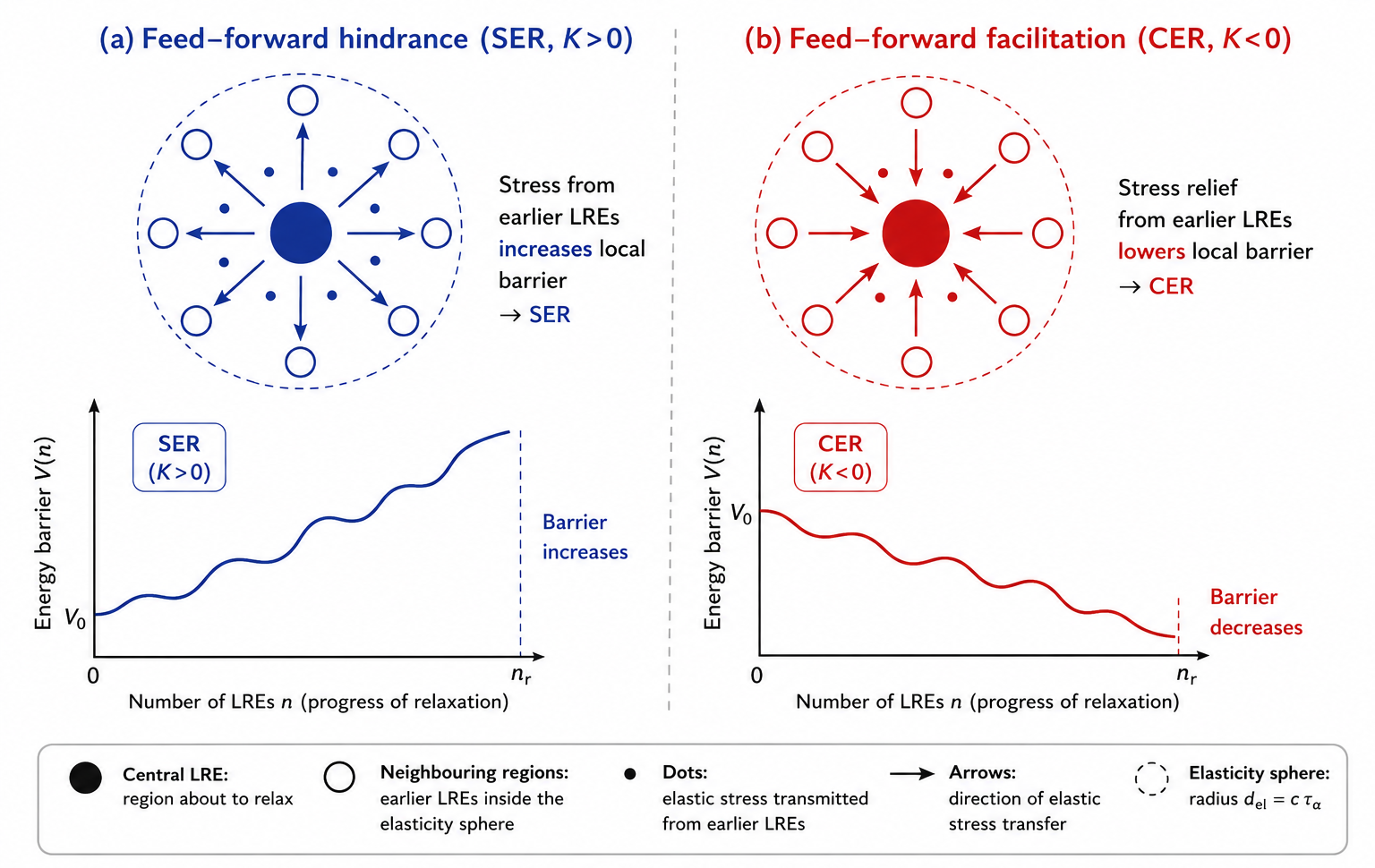}
\caption{Schematic representation of the two feedback mechanisms underlying
the Trachenko--Zaccone equation.
(a) Feed-forward hindrance in the stretched-exponential relaxation (SER)
regime. Earlier concordant local relaxation events (LREs) redistribute stress
onto regions that have not yet relaxed, progressively increasing the
activation barrier $V(n)$ for subsequent events. At the coarse-grained level
this corresponds to
$V(n)=V_0+V_1 n/n_r$, and hence to $K>0$.
(b) Facilitation in the compressed-exponential relaxation (CER) regime.
Avalanche-like rearrangements facilitate subsequent LREs, so that the
effective activation barrier decreases as relaxation proceeds. This
corresponds to
$V(n)=V_0-V_1^{\prime}n/n_r$, and hence to $K<0$.
Here $n$ is the number of LREs that have occurred and $n_r$ is their limiting
number at complete relaxation. The two apparently opposite mechanisms are
therefore described by the same nonlinear evolution equation through a change
in the sign of the feedback parameter $K$. Based on the physical framework
of Ref.~\cite{TrachenkoZaccone2021}.}
\label{fig:feedforward_LRE}
\end{figure}

The physical distinction between the two feedback mechanisms is summarized
schematically in figure~\ref{fig:feedforward_LRE}. The central quantity is the
activation barrier $V(n)$ encountered by a local relaxation event after $n$
previous LREs have occurred.

In the feed-forward regime associated with stretched-exponential relaxation
(SER), a concordant LRE locally relaxes part of the applied stress. Mechanical
equilibrium requires this stress to be redistributed to regions that have not
yet relaxed. Consequently, later LREs experience an increasingly large local
stress and therefore an increasingly large activation barrier. The dynamics
is thus self-retarding: the occurrence of earlier events progressively reduces
the probability of subsequent ones.

Deep in the glassy, strongly non-equilibrium regime, the opposite situation
can occur. A local rearrangement may trigger or facilitate neighbouring
rearrangements, giving rise to intermittent avalanche-like dynamics. In this
case, previous LREs effectively reduce the barrier encountered by subsequent
events. The dynamics is therefore self-accelerating and gives rise to
compressed-exponential relaxation (CER).

The important point illustrated by figure~\ref{fig:feedforward_LRE} is that
SER and CER correspond to opposite dependences of the same microscopic
quantity, the activation barrier, on the progress of relaxation. In the
simplest coarse-grained description these two cases can be written as

\begin{equation}
V(n)=V_0+V_1\frac{n}{n_r},
\qquad V_1>0,
\label{eq:barrier_SER}
\end{equation}

for feed-forward hindrance, and

\begin{equation}
V(n)=V_0-V_1^{\prime}\frac{n}{n_r},
\qquad V_1^{\prime}>0,
\label{eq:barrier_CER}
\end{equation}

for avalanche-like facilitation.

This sign reversal has a direct dynamical consequence because the probability
of an activated event depends exponentially on the barrier,

\begin{equation}
P_{\rm LRE}(n)
\propto
\exp\left[-\frac{V(n)}{k_{\rm B}T}\right].
\label{eq:LRE_Arrhenius}
\end{equation}

Thus, when $V(n)$ increases with $n$, each successive event becomes less
probable and the relaxation progressively slows down. When $V(n)$ decreases
with $n$, successive events become increasingly probable and the relaxation
accelerates. The distinction between stretched and compressed relaxation is
therefore encoded microscopically in the sign of the feedback between
previous LREs and the activation barrier for future LREs.

This observation is the key step leading to the Trachenko--Zaccone equation.
Rather than introducing the stretching or compression exponent $\beta$ at the
level of an assumed relaxation function, the theory introduces a microscopic
feedback parameter through the evolution of the activation barrier. The
non-Debye form of the relaxation then emerges from the resulting nonlinear
rate equation.

\section{Mathematical structure of the TZ equation}
\label{sec:math}

The microscopic derivation discussed in the previous section leads to a
remarkably compact nonlinear evolution equation,

\begin{equation}
\frac{d y}{d t}
=
-\frac{y}{\tau}\exp(Ky),
\label{eq:TZ_math}
\end{equation}

where $y(t)$ is the normalized amount of the initial perturbation that remains
unrelaxed, $\tau$ sets the characteristic time scale, and $K$ is the
dimensionless feedback parameter.

Despite its simple appearance, equation~(\ref{eq:TZ_math}) contains a rich
range of dynamical behaviour. The central mathematical feature is that the
relaxation rate is not constant, as in Debye theory, but depends exponentially
on the instantaneous state of the system. The equation can therefore be
written as

\begin{equation}
\frac{d y}{d t}
=
-\Gamma(y)y,
\label{eq:TZ_Gamma}
\end{equation}

where

\begin{equation}
\Gamma(y)
=
\frac{1}{\tau}\exp(Ky).
\label{eq:Gamma_math}
\end{equation}

Thus, the entire departure from Debye relaxation is encoded in the
state-dependent rate $\Gamma(y)$.

It is convenient to introduce dimensionless time

\begin{equation}
s=\frac{t}{\tau}.
\label{eq:dimensionless_time}
\end{equation}

Equation~(\ref{eq:TZ_math}) then becomes

\begin{equation}
\frac{d y}{d s}
=
-y\exp(Ky).
\label{eq:TZ_dimensionless}
\end{equation}

Consequently, after rescaling time by $\tau$, the shape of the relaxation
curve is controlled by a single dimensionless parameter, $K$. This
one-parameter structure is one of the most useful mathematical properties of
the TZ equation.

\subsection{Exact implicit solution}
\label{subsec:exact_solution}

Equation~(\ref{eq:TZ_dimensionless}) is separable. Rearranging gives

\begin{equation}
\frac{\exp(-Ky)}{y}\,d y
=
-d s.
\label{eq:TZ_separated}
\end{equation}

The integral on the left-hand side is expressed in terms of the exponential
integral function ${\rm Ei}(x)$, defined through

\begin{equation}
\frac{d}{d x}{\rm Ei}(x)
=
\frac{\exp(x)}{x}.
\end{equation}

Integration therefore gives

\begin{equation}
{\rm Ei}(-Ky)
=
-s+C,
\label{eq:TZ_Ei_general}
\end{equation}

where $C$ is an integration constant.

For the normalized initial condition

\begin{equation}
y(0)=1,
\label{eq:TZ_initial}
\end{equation}

one obtains

\begin{equation}
C={\rm Ei}(-K),
\end{equation}

and hence

\begin{equation}
{\rm Ei}[-Ky(t)]
=
{\rm Ei}(-K)
-
\frac{t}{\tau}.
\label{eq:TZ_exact_implicit}
\end{equation}

Equation~(\ref{eq:TZ_exact_implicit}) is the exact implicit solution of the
TZ equation for $K\neq 0$. Formally, it may also be written using the inverse
exponential integral as

\begin{equation}
y(t)
=
-\frac{1}{K}
{\rm Ei}^{-1}
\left[
{\rm Ei}(-K)-\frac{t}{\tau}
\right].
\label{eq:TZ_inverse_Ei}
\end{equation}

In practice, numerical integration of equation~(\ref{eq:TZ_math}) is often
more convenient. Nevertheless, equation~(\ref{eq:TZ_exact_implicit}) is
important because it shows that the nonlinear dynamics has a closed analytic
representation and is not merely a numerical interpolation scheme.

\subsection{The Debye limit}
\label{subsec:math_Debye}

The simplest limit is obtained for

\begin{equation}
K=0.
\end{equation}

Equation~(\ref{eq:TZ_math}) then reduces exactly to

\begin{equation}
\frac{d y}{d t}
=
-\frac{y}{\tau},
\end{equation}

with solution

\begin{equation}
y(t)
=
\exp\left(-\frac{t}{\tau}\right).
\label{eq:TZ_Debye}
\end{equation}

Thus, ordinary Debye relaxation is contained exactly within the TZ equation
as its zero-feedback limit. The parameter $K$ therefore measures the
departure from exponential relaxation.

\subsection{Instantaneous relaxation rate and the sign of the feedback}
\label{subsec:instantaneous_rate}

The physical meaning of $K$ becomes especially transparent by considering
the instantaneous logarithmic decay rate,

\begin{equation}
R(t)
=
-\frac{d\ln y}{d t}.
\label{eq:R_definition}
\end{equation}

Using equation~(\ref{eq:TZ_math}) gives

\begin{equation}
R(t)
=
\frac{1}{\tau}\exp[Ky(t)].
\label{eq:R_TZ}
\end{equation}

At the beginning of the relaxation, where $y(0)=1$,

\begin{equation}
R(0)
=
\frac{1}{\tau}\exp(K).
\label{eq:R_initial}
\end{equation}

At long times, $y\rightarrow 0$, and therefore

\begin{equation}
R(t)\rightarrow\frac{1}{\tau}.
\label{eq:R_longtime}
\end{equation}

The evolution of the instantaneous rate is controlled entirely by the sign
of $K$.

For $K>0$, the initial rate is larger than $1/\tau$ and progressively
decreases toward $1/\tau$ as relaxation proceeds. This is the self-retarding
regime generated by feed-forward hindrance and corresponds to
stretched-exponential relaxation.

For $K<0$, the initial rate is smaller than $1/\tau$ and progressively
increases toward $1/\tau$. This is the self-accelerating regime associated
with facilitation and compressed-exponential relaxation.

The three regimes can therefore be summarized as

\begin{equation}
K>0:
\qquad
{\rm SER},
\qquad
\beta<1,
\label{eq:SER_regime}
\end{equation}

\begin{equation}
K=0:
\qquad
{\rm Debye},
\qquad
\beta=1,
\label{eq:Debye_regime}
\end{equation}

and

\begin{equation}
K<0:
\qquad
{\rm CER},
\qquad
\beta>1.
\label{eq:CER_regime}
\end{equation}

This sign structure is one of the central mathematical properties of the TZ
equation: stretched and compressed relaxation occupy two sides of the same
nonlinear dynamical law.

\subsection{Comparison with earlier microscopic models of SER and CER}
\label{subsec:earlier_SER_CER_models}

It is useful at this point to compare the mathematical structure of the
TZ equation with two influential microscopic approaches developed
specifically for stretched- and compressed-exponential relaxation.

On the stretched-exponential side, the diffusion-to-traps framework discussed
in section~\ref{subsec:stretched_exponential_mauro} provides characteristic
values of the KWW exponent, including $\beta=3/5$ and $\beta=3/7$ under
different assumptions concerning the active relaxation pathways. These
results provide important examples of how stretched-exponential relaxation
can emerge from a specific microscopic mechanism.

On the compressed-exponential side, an influential microscopic description
was proposed by Bouchaud and Pitard~\cite{BouchaudPitard2001}.

They considered the elastic strain fields generated by
randomly occurring local micro-collapses in soft glassy materials. In the
characteristic early-time regime, their model predicts a compressed
exponential with the specific exponent

\begin{equation}
\beta=\frac{3}{2}.
\label{eq:BouchaudPitard_beta}
\end{equation}

The two approaches are conceptually important because they demonstrate that
both stretched- and compressed-exponential relaxation can arise from
specific microscopic mechanisms rather than merely from empirical fitting
functions. At the same time, the characteristic exponents follow from the
particular physical assumptions of the respective models: effective
diffusion dimensionality in the Phillips description and elastic
micro-collapse dynamics in the Bouchaud--Pitard theory.

The mathematical perspective of the TZ equation is different. Instead of
selecting particular characteristic values of $\beta$, it introduces a
continuous nonlinear feedback parameter $K$. Variation of this single
parameter generates a continuous family of relaxation curves,

\begin{equation}
K>0
\quad\longrightarrow\quad
0<\beta<1,
\label{eq:TZ_SER_continuum}
\end{equation}

\begin{equation}
K=0
\quad\longrightarrow\quad
\beta=1,
\label{eq:TZ_Debye_point}
\end{equation}

and

\begin{equation}
K<0
\quad\longrightarrow\quad
\beta>1.
\label{eq:TZ_CER_continuum}
\end{equation}

Thus, the values $\beta=3/5$ and $\beta=3/7$ characteristic of the
diffusion-to-traps description, as well as $\beta=3/2$ characteristic of
the early-time compressed regime of the Bouchaud--Pitard model, correspond
within the TZ framework to particular values of a more general continuous
mapping $\beta(K)$. The principal distinction is therefore not simply that
the TZ equation reproduces SER or CER, but that the two regimes are connected
continuously through the Debye point $(K,\beta)=(0,1)$ within the same
first-order nonlinear evolution equation.

\subsection{Relation to the Kohlrausch--Williams--Watts law}
\label{subsec:KWW_mapping}

The comparison with earlier microscopic models naturally raises the question
of how the continuous feedback parameter $K$ of the TZ equation is related
quantitatively to the conventional phenomenological measure of non-Debye
relaxation, namely the Kohlrausch--Williams--Watts exponent $\beta$.
The KWW relaxation function is

\begin{equation}
y_{\rm KWW}(t)
=
\exp
\left[
-\left(\frac{t}{\tau_{\rm KWW}}\right)^\beta
\right].
\label{eq:KWW_math}
\end{equation}

The exponent $\beta$ classifies the relaxation as stretched when
$0<\beta<1$, exponential when $\beta=1$, and compressed when $\beta>1$.

Beyond the diffusion-to-traps and elastic micro-collapse pictures discussed
above, a number of microscopic and mesoscopic approaches have been proposed
to rationalize non-Debye relaxation. These include descriptions based on the
vibrational density of states and boson-peak modes
\cite{CuiMilkusZaccone2017PLA,CuiMilkusZaccone2017PRE}, atomistic simulations
of metallic-glass-forming liquids displaying both stretched and compressed
relaxation \cite{WuKobWangXu2018}, free-volume descriptions of strained
metallic glasses \cite{ChenSunWang2024}, and recent approaches emphasizing
intermittent cluster dynamics and anomalous transport
\cite{RiechersDasDufresneDerletMaass2024}. A broader discussion of relaxation
mechanisms in disordered solids can be found in Ref.~\cite{Zaccone2020},
while recent perspectives emphasize the rich hierarchy of structural
dynamics and transport processes observed in metallic glasses
\cite{DerletRiechersMaass2026}.

The conceptual difference between equations~(\ref{eq:TZ_math}) and
(\ref{eq:KWW_math}) is important. In the KWW description, $\beta$ is
introduced directly as a phenomenological shape parameter. In the TZ
description, by contrast, the fundamental parameter is $K$, which controls
the nonlinear feedback in the rate equation. The apparent KWW exponent
$\beta$ emerges only after solving the dynamical equation and fitting the
resulting relaxation curve.

Numerical solutions of the TZ equation show that this procedure reproduces
both stretched- and compressed-exponential relaxation over the physically
relevant time window. Increasing positive $K$ generates progressively
stronger stretching, whereas increasingly negative $K$ generates stronger
compression.


\begin{figure}[ht]
\centering
\includegraphics[width=0.82\textwidth]
{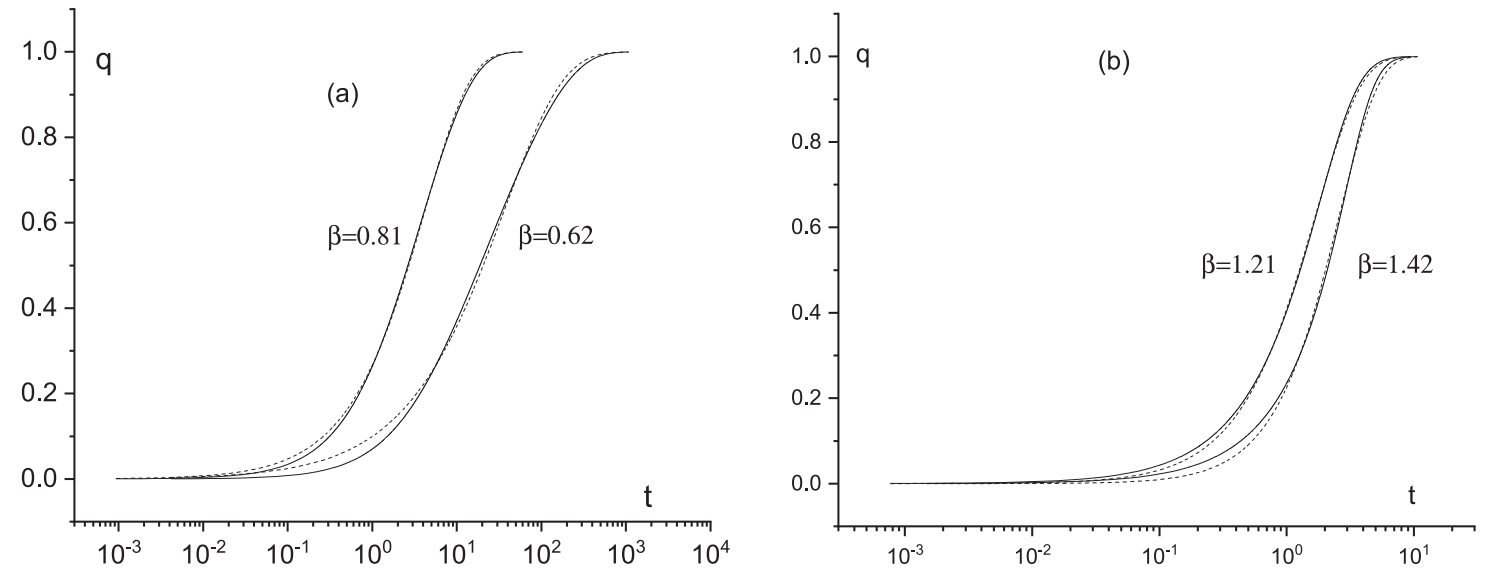}
\caption{Numerical solutions illustrating the two non-Debye regimes generated
by the Trachenko--Zaccone equation. For positive feedback parameter $K>0$,
the numerical solutions are accurately represented by stretched-exponential
KWW functions with $\beta<1$ (SER). For negative $K<0$, the same evolution
equation generates compressed-exponential relaxation with $\beta>1$ (CER).
The Debye law, $\beta=1$, lies between these two regimes and corresponds to
$K=0$. The figure illustrates that stretched and compressed relaxation do not
require separate phenomenological equations but emerge from opposite signs
of the same nonlinear feedback parameter.}
\label{fig:TZ_SER_CER_solutions}
\end{figure}

Figure~\ref{fig:TZ_SER_CER_solutions} illustrates this result directly. The
same first-order nonlinear equation produces relaxation curves that are well
represented by KWW functions on either side of $\beta=1$. The parameter $K$
therefore provides a continuous dynamical coordinate connecting SER, Debye
relaxation and CER.

\subsection{Mapping between the nonlinear parameter $K$ and the KWW
exponent $\beta$}
\label{subsec:beta_K}

The relation between the microscopic or dynamical parameter $K$ and the
phenomenological KWW exponent $\beta$ can be quantified numerically. For each
value of $K$, equation~(\ref{eq:TZ_dimensionless}) is integrated with
$y(0)=1$, and the resulting relaxation curve is fitted to
equation~(\ref{eq:KWW_math}). This procedure defines a numerical mapping

\begin{equation}
\beta=\beta(K).
\label{eq:betaK_general}
\end{equation}

\begin{figure}[ht]
\centering
\includegraphics[width=0.78\textwidth]
{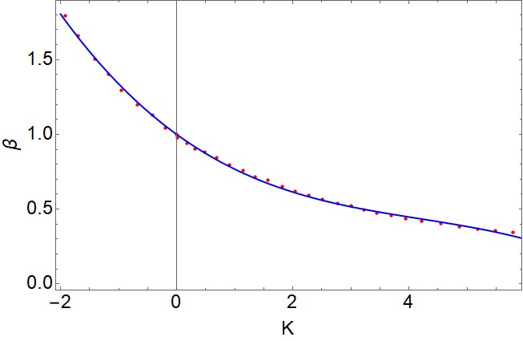}
\caption{Numerical mapping between the nonlinear feedback parameter $K$ of
the Trachenko--Zaccone equation and the effective
Kohlrausch--Williams--Watts exponent $\beta$. Each point is obtained by
numerically solving $d y/d s=-y\exp(Ky)$ with $y(0)=1$ and fitting the
resulting relaxation curve to a KWW function. The mapping passes through
$(K,\beta)=(0,1)$, corresponding to Debye relaxation. Positive $K$ produces
stretched-exponential relaxation ($\beta<1$), while negative $K$ produces
compressed-exponential relaxation ($\beta>1$). The solid curve provides a
smooth representation of the numerical mapping.}
\label{fig:beta_K_mapping}
\end{figure}

The resulting mapping is shown in figure~\ref{fig:beta_K_mapping}. It is
continuous and monotonic over the range considered. In particular,

\begin{equation}
\beta(0)=1,
\end{equation}

as required by the exact Debye limit.

Positive $K$ corresponds to $\beta<1$, whereas negative $K$ corresponds to
$\beta>1$. The mapping therefore translates the empirical KWW exponent into
the nonlinear feedback parameter of the TZ equation.


Over the numerical range explored in Ref.~\cite{TrachenkoZaccone2021}, the
mapping can be represented approximately by a low-order polynomial,

\begin{equation}
\beta
\simeq
1
-0.280831K
+0.0523623K^2
-0.00416614K^3.
\label{eq:betaK_polynomial}
\end{equation}

An alternative positive-definite representation is

\begin{equation}
\beta
\simeq
\exp\left(
-0.262807K
+0.017930K^2
-0.00092312K^3
\right).
\label{eq:beta_K_exponential}
\end{equation}

These expressions should be understood as convenient numerical
parameterizations of the mapping rather than as independent theoretical
laws. The fundamental dynamical quantity in the TZ theory remains $K$.

This distinction is conceptually useful. Whereas $\beta$ characterizes the
shape of an observed relaxation curve, $K$ quantifies the nonlinear feedback
that generates that shape. The correspondence $\beta(K)$ therefore provides
the bridge between the conventional phenomenology of KWW relaxation and the
underlying nonlinear dynamics.

\subsection{Early- and late-time behaviour}
\label{subsec:asymptotics}

Further insight follows directly from equation~(\ref{eq:TZ_math}). At very
short times, $y$ remains close to its initial value $y=1$. The initial slope
is therefore

\begin{equation}
\left.
\frac{d y}{d t}
\right|_{t=0}
=
-\frac{\exp(K)}{\tau}.
\label{eq:initial_slope}
\end{equation}

Thus, positive $K$ produces a comparatively rapid initial decay followed by
progressive slowing, whereas negative $K$ produces a slow initial response
followed by acceleration.

At long times, $y\ll 1$, and the exponential can be expanded as

\begin{equation}
\exp(Ky)
=
1+Ky+\frac{K^2y^2}{2}+\cdots .
\label{eq:expansion_longtime}
\end{equation}

The TZ equation consequently becomes

\begin{equation}
\frac{d y}{d t}
=
-\frac{y}{\tau}
-\frac{K}{\tau}y^2
+O(y^3).
\label{eq:TZ_longtime_expansion}
\end{equation}

To leading order,

\begin{equation}
\frac{d y}{d t}
\simeq
-\frac{y}{\tau},
\end{equation}

so that the asymptotic tail approaches ordinary exponential decay,

\begin{equation}
y(t)\sim C\exp(-t/\tau),
\qquad
t\rightarrow\infty,
\label{eq:TZ_asymptotic_tail}
\end{equation}

where $C$ is a constant determined by the preceding nonlinear evolution.

This result is mathematically important. The KWW form is an extremely
accurate effective representation of the nonlinear relaxation over the
experimentally relevant time window, but the TZ equation is not identical to
the KWW function at all times. Its ultimate asymptotic tail is exponential.
The stretched or compressed behaviour emerges from the intermediate-time
evolution of the state-dependent rate.

\subsection{Temperature dependence and the SER--CER crossover}
\label{subsec:T_dependence}

In the microscopic derivation, the feedback parameter is related to the
change of activation barrier through

\begin{equation}
K=\frac{V_1}{k_{\rm B}T}
\label{eq:K_temperature}
\end{equation}

for the feed-forward regime. Since $V_1$ itself depends on the elasticity
length and hence on the structural relaxation time, $K$ is generally
temperature dependent. The effective KWW exponent is therefore also
temperature dependent through the mapping

\begin{equation}
\beta(T)=\beta[K(T)].
\label{eq:beta_temperature}
\end{equation}

For systems in which the relaxation mechanism changes across the glass
transition, one may introduce a temperature-dependent fraction $f(T)$ of
facilitating, avalanche-like processes. The effective activation barrier can
then be represented schematically as

\begin{equation}
V(q,T)
=
[1-f(T)]V_{\rm SER}(q)
+
f(T)V_{\rm CER}(q),
\label{eq:mixed_barrier}
\end{equation}

with

\begin{equation}
V_{\rm SER}(q)=V_0+V_1q
\label{eq:V_SER_mix}
\end{equation}

and

\begin{equation}
V_{\rm CER}(q)=V_0-V_1q.
\label{eq:V_CER_mix}
\end{equation}

A smooth crossover may, for example, be represented by a sigmoid function of
inverse temperature,

\begin{equation}
f(T)
=
\frac{1}
{1+\exp[-A(1/T-1/T_0)]},
\label{eq:f_temperature}
\end{equation}

where $T_0$ identifies the crossover region and $A$ controls its sharpness.

This construction provides a natural route for describing a continuous
temperature-driven transition between stretched and compressed relaxation.
It is particularly relevant to metallic glasses, where experiments reveal a
sharp change of the effective KWW exponent around the glass-transition
region. We return to this comparison in the next section.

\subsection{A nonlinear generalization of Debye relaxation}
\label{subsec:nonlinear_Debye}

The mathematical structure discussed above suggests a particularly simple
interpretation of the TZ equation. Debye relaxation assumes

\begin{equation}
\Gamma={\rm constant},
\end{equation}

whereas the TZ equation replaces the constant rate by

\begin{equation}
\Gamma(y)
=
\frac{1}{\tau}\exp(Ky).
\end{equation}

In this sense, the TZ equation may be viewed as a minimal nonlinear
generalization of Debye relaxation in which the relaxing variable modifies
its own instantaneous rate.

This perspective also clarifies why the same mathematical structure can
appear in systems whose microscopic physics is very different from that of
supercooled liquids. What is essential mathematically is not the particular
origin of the activation barrier, but the existence of a feedback mechanism
through which the instantaneous state modifies the rate of subsequent
evolution.

The hierarchy of relaxation regimes can therefore be summarized as

\begin{equation}
K>0
\quad\longrightarrow\quad
\beta<1
\quad\longrightarrow\quad
{\rm stretched\ relaxation},
\end{equation}

\begin{equation}
K=0
\quad\longrightarrow\quad
\beta=1
\quad\longrightarrow\quad
{\rm Debye\ relaxation},
\end{equation}

and

\begin{equation}
K<0
\quad\longrightarrow\quad
\beta>1
\quad\longrightarrow\quad
{\rm compressed\ relaxation}.
\end{equation}

The usefulness of the TZ equation therefore lies not merely in reproducing
non-Debye relaxation curves. Its more fundamental feature is that a single
first-order nonlinear differential equation continuously connects slowing,
linear and accelerating dynamical regimes through one feedback parameter.
This structure will become particularly important in the applications
discussed in the following section.

\section{Relaxation spectra and broader non-Debye frameworks}
\label{sec:relaxation_spectra}

The nonlinear relaxation framework developed above raises a more general
issue. A single relaxation time is rarely sufficient to describe the dynamics
of a complex material. Even when the elementary microscopic process is
Debye-like, the experimentally observed response may contain a large number,
or even a continuous distribution, of characteristic times. This observation
provides an important complementary perspective on the non-Debye phenomenology
discussed throughout this review. 

The historical Debye model was developed for the orientational polarization
of polar molecules in simple media \cite{Debye1929}. Its frequency-domain
response is characterized by a single relaxation time. The same mathematical
structure appears in mechanics in the Maxwell element, in which a spring and
dashpot in series produce a single exponential stress relaxation. This
one-to-one analogy is useful, but it becomes increasingly incomplete for
polymeric and glassy materials.

Even an ideal unentangled polymer chain contains a hierarchy of normal modes
rather than a single relaxation time. Rouse theory gives a spectrum of
exponential modes with characteristic times scaling approximately as
$\tau_p\propto p^{-2}$, while Zimm theory modifies this spectrum through
hydrodynamic interactions \cite{Rouse1953,Zimm1956,DoiEdwards1986,
RubinsteinColby2003}. Entangled polymers introduce further processes,
including reptation, contour-length fluctuations and constraint release
\cite{deGennes1971,DoiEdwards1986,McLeish2002Tube,
LikhtmanMcLeish2002}. Physical aging, secondary relaxations and structural
heterogeneity broaden the dynamical spectrum still further
\cite{McCrum1967,Ferry1980,Struik1978,WardSweeney2013}.

\subsection{From a single relaxation time to a relaxation spectrum}
\label{subsec:relaxation_spectrum}

In the most general linear-response description, a normalized relaxation
function may be represented as a superposition of Debye modes,

\begin{equation}
\phi(t)
=
\int_0^{\infty}
g(\tau)
\exp\left(-\frac{t}{\tau}\right)
\dd\tau,
\label{eq:general_relaxation_spectrum}
\end{equation}

where $g(\tau)$ is a normalized distribution of relaxation times,

\begin{equation}
\int_0^{\infty}g(\tau)\dd\tau=1.
\label{eq:spectrum_normalization}
\end{equation}

The corresponding frequency-domain susceptibility can be written as

\begin{equation}
\chi^{*}(\omega)
=
\int_0^{\infty}
\frac{g(\tau)}
{1+\ii\omega\tau}
\dd\tau,
\label{eq:spectral_susceptibility}
\end{equation}

where the precise sign of the imaginary component depends on the Fourier
convention employed.

A single Debye relaxation corresponds to

\begin{equation}
g(\tau)
=
\delta(\tau-\tau_0).
\label{eq:debye_delta}
\end{equation}

Broad or asymmetric distributions of relaxation times instead generate
non-Debye response. This simple observation provides a useful phenomenological
interpretation of many of the empirical relaxation laws used in dielectric
spectroscopy and polymer rheology.

\begin{figure}[ht]
\centering
\includegraphics[width=0.82\textwidth]
{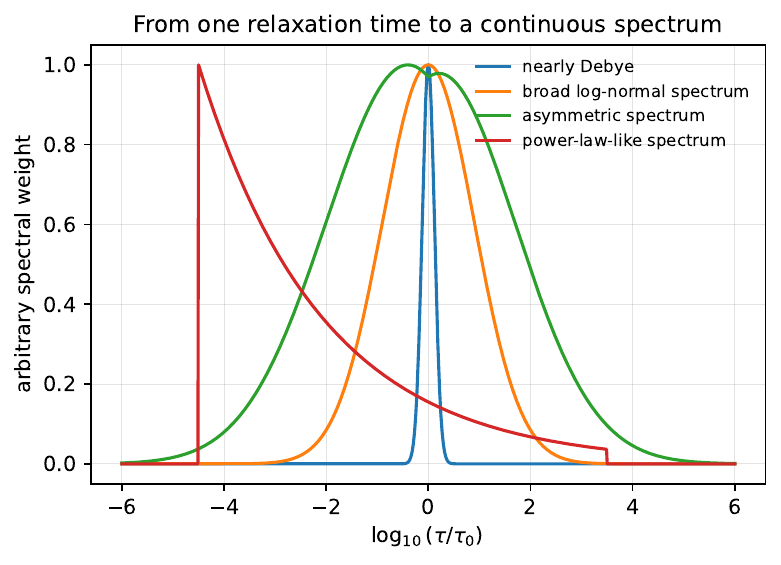}
\caption{Schematic evolution from a nearly single relaxation time to broad,
asymmetric and power-law-like distributions of relaxation times. A Debye
process corresponds to a sharply localized spectrum, whereas polymeric and
glassy materials generally exhibit broad distributions extending over many
time scales. The figure is intended as a qualitative illustration of spectral
broadening rather than as a unique inversion of a particular experimental
response.}
\label{fig:relaxation_time_spectra}
\end{figure}

Figure~\ref{fig:relaxation_time_spectra} emphasizes that non-Debye relaxation
does not necessarily imply that each microscopic process is itself
non-exponential. A broad macroscopic response may also result from a
superposition of many elementary exponential modes. This interpretation is
particularly natural in polymer physics, where a hierarchy of molecular
relaxation modes is already present at the level of chain dynamics.

\subsection{The Debye and Maxwell limits}
\label{subsec:debye_maxwell}

For a single relaxation variable $x(t)$ approaching equilibrium, Debye
relaxation follows from

\begin{equation}
\frac{\dd x}{\dd t}
=
-\frac{x-x_{\rm eq}}{\tau}.
\label{eq:debye_general}
\end{equation}

Following removal of a perturbation, the normalized relaxation is

\begin{equation}
\phi_{\rm D}(t)
=
\exp\left(-\frac{t}{\tau}\right).
\label{eq:debye_phi}
\end{equation}

For the dielectric convention

\begin{equation}
\epsilon^{*}(\omega)
=
\epsilon'(\omega)
-
\ii\epsilon''(\omega),
\end{equation}

the Debye response is

\begin{equation}
\epsilon^{*}(\omega)
=
\epsilon_{\infty}
+
\frac{\Delta\epsilon}
{1+\ii\omega\tau},
\label{eq:debye_frequency}
\end{equation}

where

\begin{equation}
\Delta\epsilon
=
\epsilon_s-\epsilon_{\infty}.
\end{equation}

Its real and loss components are

\begin{equation}
\epsilon'(\omega)
=
\epsilon_{\infty}
+
\frac{\Delta\epsilon}
{1+\omega^2\tau^2},
\label{eq:debye_real}
\end{equation}

and

\begin{equation}
\epsilon''(\omega)
=
\frac{\Delta\epsilon\,\omega\tau}
{1+\omega^2\tau^2}.
\label{eq:debye_loss}
\end{equation}

The maximum of the loss peak occurs at $\omega\tau=1$.

The mechanical analogue is the Maxwell element. A spring of modulus $G_0$
in series with a dashpot of viscosity $\eta$ gives

\begin{equation}
G(t)
=
G_0
\exp\left(-\frac{t}{\tau_M}\right),
\qquad
\tau_M=\frac{\eta}{G_0}.
\label{eq:maxwell_time}
\end{equation}

Its complex modulus is

\begin{equation}
G^{*}(\omega)
=
G_0
\frac{\ii\omega\tau_M}
{1+\ii\omega\tau_M}.
\label{eq:maxwell_complex}
\end{equation}

Thus Debye dielectric relaxation and the single Maxwell element represent
the corresponding single-time-scale limits of dielectric and mechanical
linear response.

\begin{figure}[ht]
\centering
\includegraphics[width=0.86\textwidth]
{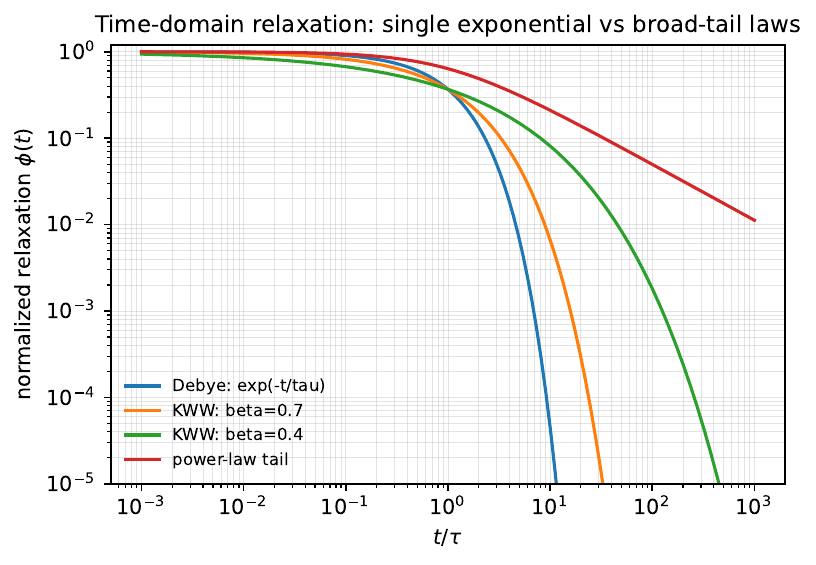}
\caption{Time-domain comparison between a single Debye exponential,
stretched-exponential KWW relaxation with two representative values of
$\beta$, and a power-law-tailed relaxation. Decreasing $\beta$ produces a
progressively broader decay in time, while a power-law tail extends over an
even wider temporal window.}
\label{fig:time_domain_decays}
\end{figure}

\subsection{Cole--Cole, Cole--Davidson and Havriliak--Negami laws}
\label{subsec:frequency_domain_laws}

Several compact empirical functions have been developed to represent the
broadening and asymmetry of experimentally observed loss spectra. Their
importance is particularly evident in polymers and glass-forming liquids.

The Cole--Cole expression is

\begin{equation}
\chi^{*}_{\rm CC}(\omega)
=
\frac{\Delta\chi}
{1+(\ii\omega\tau_0)^{\alpha}},
\qquad
0<\alpha\leq1.
\label{eq:cole_cole}
\end{equation}

The Debye limit is recovered for $\alpha=1$. Decreasing $\alpha$ produces a
broader and approximately symmetric dispersion on a logarithmic frequency
scale \cite{ColeCole1941}.

The Cole--Davidson form is

\begin{equation}
\chi^{*}_{\rm DC}(\omega)
=
\frac{\Delta\chi}
{(1+\ii\omega\tau_0)^{\beta}},
\qquad
0<\beta\leq1.
\label{eq:davidson_cole}
\end{equation}

In contrast to the Cole--Cole law, the Cole--Davidson function generates an
asymmetric response and a modified high-frequency wing
\cite{DavidsonCole1951}.

A still more general representation is the Havriliak--Negami form,

\begin{equation}
\chi^{*}_{\rm HN}(\omega)
=
\frac{\Delta\chi}
{\left[
1+(\ii\omega\tau_0)^{\alpha}
\right]^{\beta}},
\qquad
0<\alpha\leq1,
\qquad
0<\alpha\beta\leq1.
\label{eq:havriliak_negami}
\end{equation}

It contains the previous laws as limiting cases,

\begin{equation}
\alpha=\beta=1
\quad\Longrightarrow\quad
{\rm Debye},
\label{eq:HN_Debye}
\end{equation}

\begin{equation}
\beta=1
\quad\Longrightarrow\quad
{\rm Cole\!-\!Cole},
\label{eq:HN_CC}
\end{equation}

and

\begin{equation}
\alpha=1
\quad\Longrightarrow\quad
{\rm Cole\!-\!Davidson}.
\label{eq:HN_DC}
\end{equation}

This two-parameter flexibility explains the widespread use of the
Havriliak--Negami law in broadband dielectric spectroscopy of polymers and
glasses \cite{HavriliakNegami1967}.

\begin{figure}[ht]
\centering
\includegraphics[width=0.90\textwidth]
{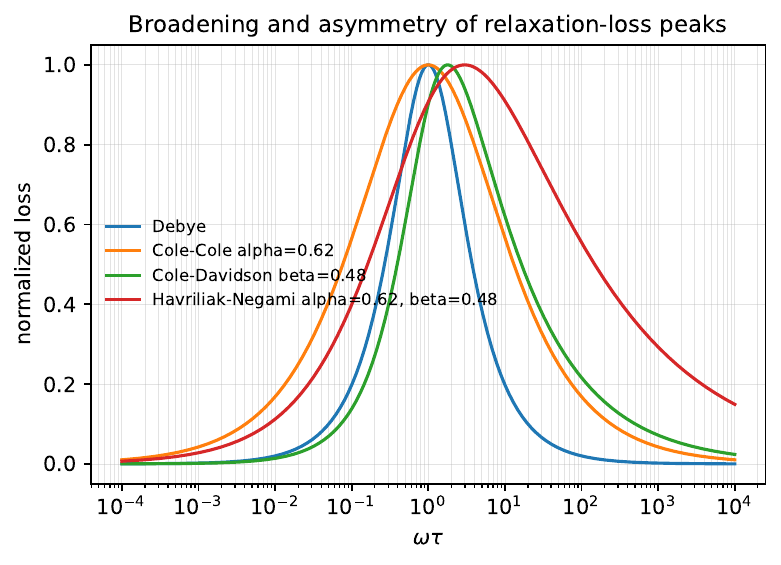}
\caption{Normalized frequency-domain loss peaks for Debye, Cole--Cole,
Cole--Davidson and Havriliak--Negami response functions. Cole--Cole
broadening is approximately symmetric on the logarithmic frequency scale,
whereas Cole--Davidson and Havriliak--Negami response functions generate
asymmetric loss peaks and extended high-frequency wings. The curves are
schematic and use representative non-Debye parameters.}
\label{fig:frequency_loss_laws}
\end{figure}

The different laws are particularly transparent in the complex plane.
For a single Debye mode the normalized susceptibility traces a semicircle.
Broad distributions of relaxation times depress and distort this arc.

\begin{figure}[ht]
\centering
\includegraphics[width=0.70\textwidth]
{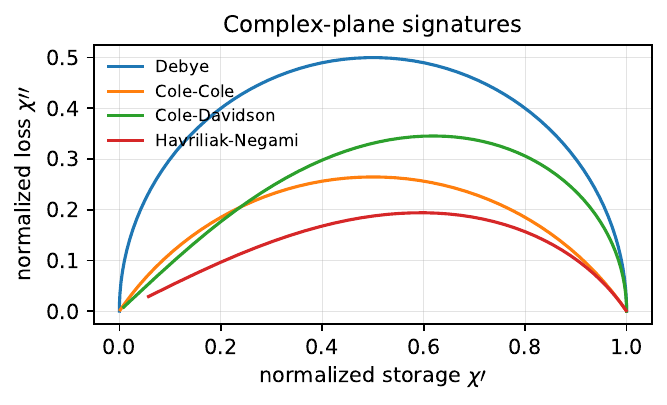}
\caption{Complex-plane signatures of Debye and common non-Debye relaxation
laws. A single Debye process produces the characteristic semicircle.
Cole--Cole broadening depresses the arc, whereas Cole--Davidson and
Havriliak--Negami relaxation introduce asymmetric distortions associated
with broader and skewed relaxation-time spectra.}
\label{fig:complex_plane_laws}
\end{figure}

\subsection{Relation to KWW relaxation}
\label{subsec:spectra_KWW}

The Kohlrausch--Williams--Watts function discussed earlier in this review is
a time-domain representation,

\begin{equation}
\phi_{\rm KWW}(t)
=
\exp
\left[
-\left(
\frac{t}{\tau_K}
\right)^{\beta_K}
\right],
\qquad
0<\beta_K\leq1.
\label{eq:KWW_spectrum_section}
\end{equation}

It reduces to Debye relaxation at $\beta_K=1$, whereas decreasing $\beta_K$
corresponds to an increasingly broad effective distribution of relaxation
times.

For general $\beta_K$, the Fourier transform of the KWW function does not
have a simple elementary form. Havriliak--Negami functions are consequently
often used as compact frequency-domain representations of KWW-like
relaxation. The mapping between KWW and Havriliak--Negami parameters is not
exact and depends on the fitting criterion and frequency interval
\cite{LindseyPatterson1980,Alvarez1991,AlvarezAlegriaColmenero1993}.

This point is relevant to the TZ equation. The KWW exponent $\beta$ obtained
from a TZ relaxation curve characterizes the shape of that curve, but it
does not imply that the underlying microscopic dynamics consists of a unique
distribution of independent Debye modes. In the original TZ derivation,
non-Debye behaviour arises dynamically because the relaxation rate changes
with the instantaneous state. A representation in terms of a relaxation-time
spectrum and a representation in terms of a nonlinear state-dependent rate
are therefore complementary mathematical descriptions rather than identical
microscopic statements.

\subsection{Fractional and power-law descriptions}
\label{subsec:fractional_powerlaw}

Broad relaxation spectra may also be represented using fractional
differential equations. A fractional Debye equation can be written
schematically as

\begin{equation}
D_t^{\alpha}\phi(t)
=
-\tau^{-\alpha}\phi(t),
\qquad
0<\alpha\leq1,
\label{eq:fractional_debye}
\end{equation}

with the Mittag--Leffler solution

\begin{equation}
\phi(t)
=
E_{\alpha}
\left[
-\left(
\frac{t}{\tau}
\right)^{\alpha}
\right].
\label{eq:mittag_leffler}
\end{equation}

Fractional viscoelastic models similarly replace ordinary spring or dashpot
elements by spring-pot elements whose complex modulus scales as

\begin{equation}
G^{*}(\omega)
\propto
(\ii\omega)^{\alpha}.
\label{eq:fractional_modulus}
\end{equation}

Such formulations provide economical representations of broad power-law
memory and can replace large Prony-series expansions in polymer rheology
\cite{BagleyTorvik1983,Koeller1984,Mainardi2010}. Fractional differential
equations have developed into a broad mathematical and constitutive framework
for describing anomalous transport, power-law rheology and multiscale
viscoelastic response
\cite{ZayernouriWangShenKarniadakis2024,
JaishankarMcKinley2013,
JaishankarMcKinley2014,
RathinarajMcKinleyKeshavarz2021,
SongHoltenAndersenMcKinley2023,
SuzukiGulianZayernouriDElia2023}.

Applications now span systems ranging from complex fluids and biological
materials to polymers, elastomers and hydrogels. Examples include fractional
descriptions of single-cell rheology, liquid foods, visco-elasto-plastic
materials, polyurea elastomers and nanocomposites, and ionically
cross-linked polymer networks
\cite{DasWaeterloosClasenMcKinley2025,
WagnerBarbatiEngmannBurbidgeMcKinley2017,
TzelepisEtAl2023IPDI,
SuzukiEtAl2016Fractional,
TzelepisKhoshnevisZayernouriGinzburg2023,
DuttaGinzburgVasilyevZussman2026}.

Related power-law response occurs in disordered dielectric materials. The
frequency-dependent conductivity is often represented empirically as

\begin{equation}
\sigma'(\omega)
=
\sigma_{\rm dc}
+
A\omega^{s},
\qquad
0<s<1,
\label{eq:jonscher}
\end{equation}

as in the universal dielectric response of Jonscher
\cite{Jonscher1977,NgaiJonscherWhite1979}.

These approaches emphasize a point that is also relevant to future
generalizations of the TZ framework. A broad experimental relaxation law can
encode several distinct kinds of physics: a distribution of elementary
relaxation times, a non-local memory kernel, fractional dynamics, or a
state-dependent nonlinear rate. Distinguishing between these interpretations
requires information beyond the quality of a fit to a single relaxation
curve.

The TZ equation occupies a distinctive position within this broader
landscape. It achieves non-Debye relaxation without introducing a relaxation
spectrum or a fractional operator explicitly. Instead, the instantaneous
relaxation rate changes dynamically with the state variable itself. A natural
future direction is therefore to combine this nonlinear rate-feedback
mechanism with the spectral and memory effects described above, for example
through distributions of $\tau$ and $K$, coupled TZ modes, or fractional and
non-local extensions of the TZ evolution operator.

The comparison above highlights an important distinction that will recur in the following sections. A broad relaxation spectrum and a nonlinear state-dependent relaxation rate are two different mechanisms for generating non-Debye response. In the former, the macroscopic relaxation is constructed from a superposition of modes with different characteristic times. In the latter, as in the TZ equation, even a single relaxation variable evolves non-exponentially because its instantaneous rate changes with its state. These descriptions need not be mutually exclusive. In heterogeneous materials such as polymer and structural glasses, a natural generalization is therefore to associate a TZ-type nonlinear evolution law with each member of a distribution of relaxation modes. The resulting theory would contain both relaxation-time polydispersity and nonlinear feedback, thereby connecting the spectral viewpoint developed in this section with the dynamical viewpoint underlying the TZ equation.


\section{Applications to polymer glasses: volume and enthalpy relaxation}
\label{sec:polymer_applications}

As discussed above, nonexponential relaxations are ubiquitous in nature. One class
of materials where their role is crucial is glasses (network, metallic, organic, or polymer). In
glasses, the dependence of viscosity (or characteristic time) on temperature (in isobaric,
atmospheric-pressure experiments) is non-Arrhenius
\cite{Vogel1921,Fulcher1925,TammannHesse1926,WilliamsLandelFerry1955,
AvramovMilchev1988,MauroEtAl2009,Ginzburg2020},
and the relaxation (dielectric, stress, volume, and enthalpy) is non-Debye
\cite{Kohlrausch1854,WilliamsWatts1970,AlvarezAlegriaColmenero1993,
HavriliakNegami1966,Tool1946,Narayanaswamy1971,MoynihanEtAl1976}.
Developing a fundamental understanding of those processes -- or even a self-consistent
phenomenological model -- was and remains a major challenge for researchers in the field.

One particular question, to which we will return several times, concerns the origin of non-Debye relaxation. The ``material time'' paradigm formulated by
Narayanaswami \cite{Narayanaswamy1971} and subsequently expanded by Dyre and
co-workers \cite{DouglassDyre2022,RiechersEtAl2022,BohmerEtAl2024}, implies that the
relaxation is Debye-like but only relative to the ``material time'' $\xi$, which is a nonlinear
function of the laboratory time, $t$. Thus, the enthalpy or volume relaxation equation can be
written as,

\begin{equation}
\frac{d\delta}{d\xi}+\delta=0 ,
\label{eq:polymer_1a}
\end{equation}

\begin{equation}
\xi=
\left[
\int_0^t
\frac{d t'}{\tau(t')}
\right]^\beta ,
\label{eq:polymer_1b}
\end{equation}

\begin{equation}
\ln\left[\tau(t)\right]
=
A+
\left[
\frac{x h^{*}}{RT}
+
\frac{(1-x)h^{*}}{R T_f(t)}
\right],
\label{eq:polymer_1c}
\end{equation}

\begin{equation}
\delta(t)
=
\frac{T_f(t)-T}{T}.
\label{eq:polymer_1d}
\end{equation}

Equations~(\ref{eq:polymer_1a})--(\ref{eq:polymer_1c}) represent the
well-known Tool--Narayanaswami--Moynihan (TNM) model. Here, $A$, $x$, $\beta$, and $h^{*}$ are the TNM model parameters. For more details,
see, e.g., Malek and co-workers
\cite{Malek1998,MalekEtAl2009,Malek2023}.

The ``material time'' framework naturally invites comparison with the TZ equation.
Indeed, the TZ equation,

\begin{equation}
\frac{dy}{dt}
=
-\frac{y}{\tau}
e^{Ky},
\label{eq:polymer_2a}
\end{equation}

can be re-written as,

\begin{equation}
\frac{dy}{d\xi}+y=0 ,
\label{eq:polymer_2b}
\end{equation}

With

\begin{equation}
d \xi
=
\frac{d t}{\tau}\exp(Ky).
\end{equation}

Thus, the TZ framework offers a specific functional form for the ``material time''
behavior in glasses. Within the TNM framework, the parameters $A$, $x$, $\beta$, and $h^{*}$ are often
empirical and fitted to the specific relaxation data (e.g., volume or enthalpy); the
connection to other experimental data (dielectric relaxation or mechanical measurements)
is often non-existent. The use of TZ equation reduces the number of adjustable parameters
and allows one to search for a better physical justification of those parameters.

In 2020--2022, Ginzburg and co-workers formulated the mean-field ``two-state, two-
(time)scale'' (TS2) model for the glassy dynamics and thermodynamics
\cite{Ginzburg2020,Ginzburg2021Macromolecules,Ginzburg2021SoftMatter,
Ginzburg2022ThinFilms,GinzburgZacconeCasalini2022}.
Within TS2 -- combined with the Sanchez-Lacombe
\cite{SanchezLacombe1976,LacombeSanchez1976,SanchezLacombe1978}
lattice equation of state -- the state of the
material is described by two variables, $\psi$ -- the ``solid fraction'', and $\nu$ -- the ``occupancy''
([$1-\nu$] is the fractional free volume (FFV) often used to estimate the relaxation time)
\cite{WhiteLipson2017,WhiteLipson2016,Doolittle1951,DoolittleDoolittle1957}. The
free energy of the two-state system has the form,
\begin{eqnarray}
G &=& -\epsilon^{*}\frac{r}{\nu}
\left[
\left(
\frac{\nu\psi r_S}{r}
\right)^2
\right.
\left.
+\,2\alpha_{LS}
\left(
\frac{\nu\psi r_S}{r}
\right)
\left(
\frac{\nu\{1-\psi\}r_L}{r}
\right)
\right.
\nonumber\\
&&\left.
+\,\alpha_{LL}
\left(
\frac{\nu\{1-\psi\}r_L}{r}
\right)^2
\right]
+\,k_{\rm B}T
\left[
\psi\ln
\left(
\frac{\nu\psi r_S}{r}
\right)
\right.
\nonumber\\
&&\left.
+\,\{1-\psi\}\ln
\left(
\frac{\nu\{1-\psi\}r_L}{r}
\right)
\right.
\left.
+\,r\frac{1-\nu}{\nu}
\ln(1-\nu)
\right].
\label{eq:polymer_4}
\end{eqnarray}

Here, $T$ is the absolute temperature, $k_{\rm B}$ is the Boltzmann's constant, $r_S$ and $r_L$ are the
number of lattice sites occupied by the ``Solid'' and ``Liquid'' states of the glass-former,
respectively. The van-der-Waals interaction energies are $\epsilon_{LL}$ (``Liquid''--``Liquid'' nearest
neighbors), $\epsilon_{SS}$ (``Solid''--``Solid'' nearest neighbors), and $\epsilon_{LS}$
(``Liquid''--``Solid'' nearest neighbors); we can then define
$\alpha_{LL}=\epsilon_{LL}/\epsilon_{SS}$,
$\alpha_{LS}=\epsilon_{LS}/\epsilon_{SS}$, and
$\epsilon^{*}=Z\epsilon_{SS}/2$ (where $Z$ is
the coordination number). Finally,
$r=\psi r_S+[1-\psi]r_L$. The equilibrium state of matter is
computed in the standard way by minimizing $G$ with respect to $\psi$ and $\nu$.

Importantly, it is assumed that the relaxation dynamics of the two variables are
different -- the dynamics of $\psi$ are fast (the $\beta$-process), while the dynamics of $\nu$ are slow (the
$\alpha$-process). Specifically,

\begin{equation}
\frac{d\psi}{dt}
=
\frac{\psi^{*}(\nu)-\psi}{\tau_{\beta}},
\label{eq:polymer_4a}
\end{equation}

\begin{equation}
\frac{d\nu}{dt}
=
\frac{\nu^{*}-\nu}{\tau_{\alpha}(\psi)},
\label{eq:polymer_4b}
\end{equation}

\begin{equation}
\tau_{\beta}
=
\tau_{\infty}
\exp
\left[
\frac{E_1}{RT}
\right],
\label{eq:polymer_4c}
\end{equation}

\begin{equation}
\tau_{\alpha}
=
\tau_{\infty}
\exp
\left[
\frac{E_1}{RT}
+
\frac{E_2-E_1}{RT}\psi
\right].
\label{eq:polymer_4d}
\end{equation}

Here, $E_1$ and $E_2$ are the activation energies of the ``liquid'' and ``solid'' states,
respectively, and $\tau_{\infty}$ is the ``elementary time'' of molecular processes. Also, we define
$\psi^{*}(\nu)$ as the solution of equation
$\delta G(\psi,\nu)/\delta\psi=0$, and $\nu^{*}$ as the solution of equation
$\delta G(\psi^{*}(\nu),\nu)/\delta\nu=0$. (Note that for a given temperature, $T$, $\nu^{*}$ is a single
number, and $\psi^{*}(\nu)$ is a function.)

If we consider an isothermal relaxation process, separate the small and large
timescales, and concentrate on the large timescale only, it is possible to integrate out the
fast variable $\psi$ and arrive at the following nonlinear equation for $\nu$,

\begin{eqnarray}
\frac{d\nu}{dt}
&=&
\frac{\nu^{*}-\nu}{\tau_{\alpha}(T)}
\exp
\left[
\frac{E_2-E_1}{RT}
\left(
\psi^{*}(\nu)-\psi
\right)
\right]
\nonumber \\
&&=
\frac{\nu^{*}-\nu}{\tau_{\alpha}(T)}
\exp
\left[
\frac{E_2-E_1}{RT}
\left(
\frac{d\psi^{*}}{d\nu}
\right)
(\nu^{*}-\nu)
\right].
\label{eq:polymer_5}
\end{eqnarray}

In many practical relaxation experiments, like Kovacs' up- and down-temperature
jumps \cite{Kovacs1966,Kovacs1963,Struik1997,GrassiaSimon2012,GrassiaEtAl2018},
the state of the material is changed very little (even though the characteristic
times can vary by several orders of magnitude!). Linearizing the function $\psi^{*}(\nu)$ about the equilibrium occupancy
$\nu^{*}$ gives

\begin{equation}
\frac{dy}{dt}
=
-\frac{y}{\tau_{\alpha}(T)}
\exp(Ky),
\label{eq:polymer_6a}
\end{equation}

\begin{equation}
K
=
\frac{E_2-E_1}{RT}
\left(
\frac{d\psi^{*}}{d\nu}
\right).
\label{eq:polymer_6b}
\end{equation}

Here, $\tau_{\alpha}(T)$ is the equilibrium $\alpha$-relaxation time at temperature $T$, and
$y=\nu^{*}-\nu$.
Thus, the SL-TS2 framework automatically leads to the TZ-type equation for the free volume
relaxation. However, describing the specific (not free!) volume relaxation and the enthalpy
relaxation needs additional steps, discussed below.

From the Sanchez-Lacombe framework, we can write down the expressions for the
volume, $V$, and enthalpy, $H$, as functions of $\psi$ and $\nu$,

\begin{equation}
V
=
v_{\rm sp,0}
\frac{r}{r_S\nu}
=
v_{\rm sp,0}
\frac{
\left[
\psi+
\left(
\frac{r_L}{r_S}
\right)(1-\psi)
\right]
}{\nu},
\label{eq:polymer_7a}
\end{equation}

\begin{equation}
H
\propto
-\epsilon^{*}
\frac{r}{\nu}
\left[
\left(
\frac{\nu\psi r_S}{r}
\right)^2
+
2\alpha_{LS}
\left(
\frac{\nu\psi r_S}{r}
\right)
\left(
\frac{\nu(1-\psi)r_L}{r}
\right)
+
\alpha_{LL}
\left(
\frac{\nu(1-\psi)r_L}{r}
\right)^2
\right].
\label{eq:polymer_enthalpy}
\end{equation}

Let us define the ``normalized specific volume deviation'' (NSVD), $y_V$, and the
``normalized enthalpy deviation'' (NED), $y_H$, as follows,

\begin{equation}
y_V
=
\frac{V-V^{*}}{V_I-V^{*}},
\label{eq:polymer_8a}
\end{equation}

\begin{equation}
y_H
=
\frac{H-H^{*}}{H_I-H^{*}}.
\label{eq:polymer_8b}
\end{equation}

Both NSVD and NED are functions of $\nu$, and we approximate them as linear
functions,

\begin{equation}
y_V
=
\frac{dy_V}{d\nu}
(\nu-\nu^{*}),
\label{eq:polymer_9a}
\end{equation}

\begin{equation}
y_H
=
\frac{dy_H}{d\nu}
(\nu-\nu^{*}).
\label{eq:polymer_9b}
\end{equation}

Hence, the function $\psi^{*}(\nu)$ can be
approximated as a linear one, and equation~(\ref{eq:polymer_5}) can be
re-written as,

\begin{equation}
\psi^{*}(\nu)-\psi^{*}(\nu^{*})
=
\left[
\frac{d\psi}{d\nu}
\right]_{\nu=\nu^{*}}
(\nu-\nu^{*}).
\label{eq:polymer_linear_psi}
\end{equation}

(Here and in the following, we assume that
all the derivatives are evaluated at $\nu=\nu^{*}$ and
$\psi=\psi_{\rm eq}=\psi^{*}(\nu^{*})$). Then,

\begin{equation}
\frac{dy_V}{dt}
=
-\frac{y_V}{\tau_{\alpha,\rm eq}}
\exp[Q_Vy_V],
\label{eq:polymer_10a}
\end{equation}

\begin{equation}
\frac{dy_H}{dt}
=
-\frac{y_H}{\tau_{\alpha,\rm eq}}
\exp[Q_Hy_H],
\label{eq:polymer_10b}
\end{equation}

where,

\begin{equation}
Q_V
=
-(V_I-V^{*})
\frac{E_2-E_1}{RT}
\left(
\frac{d\psi}{d\nu}
\right)
\left[
\frac{\partial V}{\partial\psi}
\frac{d\psi}{d\nu}
+
\frac{\partial V}{\partial\nu}
\right]^{-1},
\label{eq:polymer_11a}
\end{equation}

\begin{equation}
Q_H
=
-(H_I-H^{*})
\frac{E_2-E_1}{RT}
\left(
\frac{d\psi}{d\nu}
\right)
\left[
\frac{\partial H}{\partial\psi}
\frac{d\psi}{d\nu}
+
\frac{\partial H}{\partial\nu}
\right]^{-1}.
\label{eq:polymer_11b}
\end{equation}

The expressions in the square brackets on the right-hand side of
equations~(\ref{eq:polymer_11a}) and~(\ref{eq:polymer_11b}) are both negative -- as the occupancy $\nu$ is increased, the free volume, specific volume, and
enthalpy all decrease. Therefore, for the down-jump experiments (the initial temperature $T_I$
is greater than the aging temperature $T_a$), $Q_V$ and $Q_H$ are both positive, and the relaxation is
stretched-exponential \cite{RiechersEtAl2022,TrachenkoZaccone2021,NissDyreHecksher2020}.
For the up-jump experiments ($T_I < T_a$), $Q_V$ and $Q_H$ are both
negative, and the relaxation is compressed-exponential
\cite{RiechersEtAl2022,TrachenkoZaccone2021}. In the limit $T_I\rightarrow T_a$, both $Q_V$ and
$Q_H$ would approach zero, and the relaxation would become similar to the classical Debye
exponential (although, of course, the magnitude of the un-normalized volume or enthalpy
deviations would be infinitesimally small).

\begin{figure}[ht]
\centering
\includegraphics[width=0.88\textwidth]{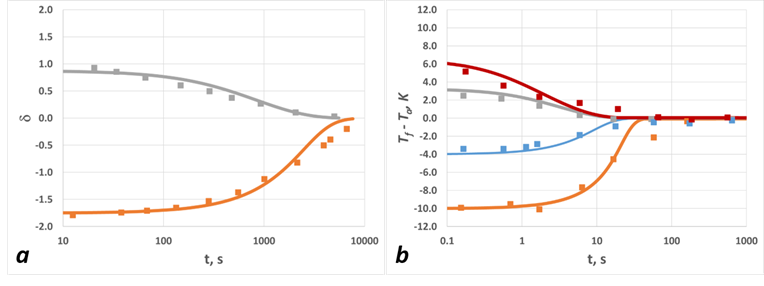}
\caption{(a) Polystyrene (PS) volume relaxation: down-jump from
110~$^\circ$C to 89~$^\circ$C (grey) and up-jump from 83~$^\circ$C
to 89~$^\circ$C (orange). The squares are the data of
Struik~\cite{Struik1966}, and the lines are the fits using the current model.
(b) Enthalpy relaxation of PS: down-jumps of 10~$^\circ$C (red) and
4~$^\circ$C (grey), and up-jumps of 10~$^\circ$C (orange) and
4~$^\circ$C (blue). In all cases, the aging temperature is
107~$^\circ$C. The squares are data from Grassia
et al.~\cite{GrassiaEtAl2018}, and the lines are the fits using the current
model.}
\label{fig:polymer_relaxation}
\end{figure}

Treating $Q_V$ and $Q_H$ as adjustable parameters, Ginzburg et al.
\cite{GinzburgGendelmanZaccone2024} re-analyzed the
volume relaxation experiments of Struik \cite{Struik1966} and the enthalpy relaxation experiments of
Grassia and co-workers \cite{GrassiaEtAl2018}. Figure~\ref{fig:polymer_relaxation}
shows the fitting results for the Struik data (a) and the
Grassia data (b). For more details, including the parameter values, see Ginzburg et al.
\cite{GinzburgGendelmanZaccone2024}.

Based on the recent ``general'' parameterization of SL-TS2
\cite{GinzburgGendelmanCasaliniZaccone2026}, it should be possible to
estimate the parameters $Q_V$ and $Q_H$ \textit{a priori}, making the model predictive instead of
descriptive. This work is ongoing.

The analysis of Ginzburg et al.~\cite{GinzburgGendelmanZaccone2024}
connects the SL--TS2 framework, the ``material time'' concept, and the
stretched or compressed exponentials via the Trachenko--Zaccone framework.
This approach allows one to successfully describe many important features of
the glassy relaxations. But is it the whole story? The limitations of this
analysis lie in the implicit ``material time'' model assumption that there is
a unique ``material time'' that can linearize (locally, not globally) any
relaxation process. Yet, it has been acknowledged by many researchers
\cite{Struik1997,GrassiaSimon2012,KovacsAklonisHutchinsonRamos1979,
RamosEtAl1988,GrassiaDAmore2011,ZhaoGrassiaSimon2021,
DAmoreCaputoGrassiaZarrelli2006,McKennaSimon2017}
that a better relaxation theory requires multiple, not single, relaxation
times. This relaxation-time polydispersity can be introduced by combining multiple
TNM-like equations, as in the KAHR model
\cite{KovacsAklonisHutchinsonRamos1979,RamosEtAl1988},
or by replacing the simple time-evolution operator $\dd y/\dd t$ with
something more complex, as discussed in
section~\ref{sec:relaxation_spectra}.

\section{Application to population dynamics}
\label{sec:population_dynamics}

Perhaps the most striking indication of the broader applicability of the
equation has come from population dynamics. A recent application showed that,
after changing the overall sign of the evolution equation from relaxation to
growth,

\begin{equation}
\frac{\dd P}{\dd t}
=
\frac{P}{\tau}\exp(KP),
\label{eq:TZ_population}
\end{equation}

Here $P$ may be understood as a normalized population variable. If the
dimensional population is used instead, $K$ has dimensions of inverse
population so that $KP$ remains dimensionless.

The same nonlinear mathematical structure can describe several distinct
regimes of global human population growth over approximately twelve
millennia \cite{ZacconeTrachenko2026}. The connection with familiar
population-growth laws follows directly from appropriate limits of the
nonlinear equation. For $K=0$, ordinary Malthusian exponential growth is
recovered. For $K<0$ and $|KP|\ll1$,

\begin{equation}
\frac{\dd P}{\dd t}
\simeq
\frac{P}{\tau}(1+KP)
=
\frac{P}{\tau}
\left(
1-\frac{P}{P_c}
\right),
\qquad
P_c=-\frac{1}{K},
\label{eq:TZ_logistic_limit}
\end{equation}

which has the logistic form to leading order. Away from this perturbative
limit, the full exponential feedback generates stretched- and
compressed-exponential growth regimes within the same nonlinear framework.

This sign change is conceptually revealing. In the relaxation problem,
$y(t)$ decreases and the two signs of $K$ generate stretched- or
compressed-exponential decay. In the population problem, the overall sign
of the dynamical equation is reversed because the relevant variable grows
rather than relaxes. The analogous nonlinear feedback therefore generates
stretched- or compressed-exponential growth. The distinction between the two
applications lies in the physical interpretation of the state variable and
the direction of evolution, whereas the nonlinear rate-feedback structure
remains unchanged.

Independent physics-based analyses of global demographic data have also
identified pronounced changes in the functional form of human population
growth. Sojecka and Drozd-Rzoska reported a transition near 1970 between
compressed-exponential and stretched-exponential growth regimes
\cite{SojeckaDrozdRzoska2024}. More recently, Yakovenko analysed global
population data extending from 10,000 BCE to the present and identified a
sequence of exponential, super-exponential and hyperbolic growth regimes,
followed in the twenty-first century by a pronounced departure from the
historical hyperbolic trend \cite{Yakovenko2025Population}. These studies
provide useful complementary examples of how concepts and mathematical
methods originating in statistical physics can reveal distinct dynamical
regimes in long-term demographic data.

In the population application of the TZ equation, however, the objective is
to place such regimes within a single nonlinear differential equation and to
associate their evolution with a continuously varying feedback parameter
rather than introducing separate growth laws for different historical epochs.

The population example should not be interpreted as implying that the
microscopic physics of glasses and demographic evolution are analogous.
They clearly are not. Rather, it illustrates a more general mathematical
point: systems with entirely different microscopic constituents may display
the same coarse-grained nonlinear dynamics when the instantaneous rate of
change depends strongly on the current state of the system. Similar
rate-feedback structures may arise in ecological systems, interacting
populations, epidemic spreading, technological adoption, network evolution
and other complex dynamical processes.


\section{Future directions}
\label{sec:future}

\subsection{Relaxation time polydispersity and nonlinearity: towards a generalized TZ equation}

An important distinction emerges when the Trachenko--Zaccone description is considered together with the conventional relaxation-spectrum picture discussed above. Non-Debye relaxation may arise from at least two conceptually distinct sources: \emph{polydispersity}, corresponding to a distribution of relaxation times, and \emph{nonlinearity}, corresponding to a state-dependent relaxation rate. These two mechanisms need not be mutually exclusive and, in realistic glassy and polymeric systems, are likely to coexist.

\begin{figure}[t]
\centering
\includegraphics[width=0.82\linewidth]{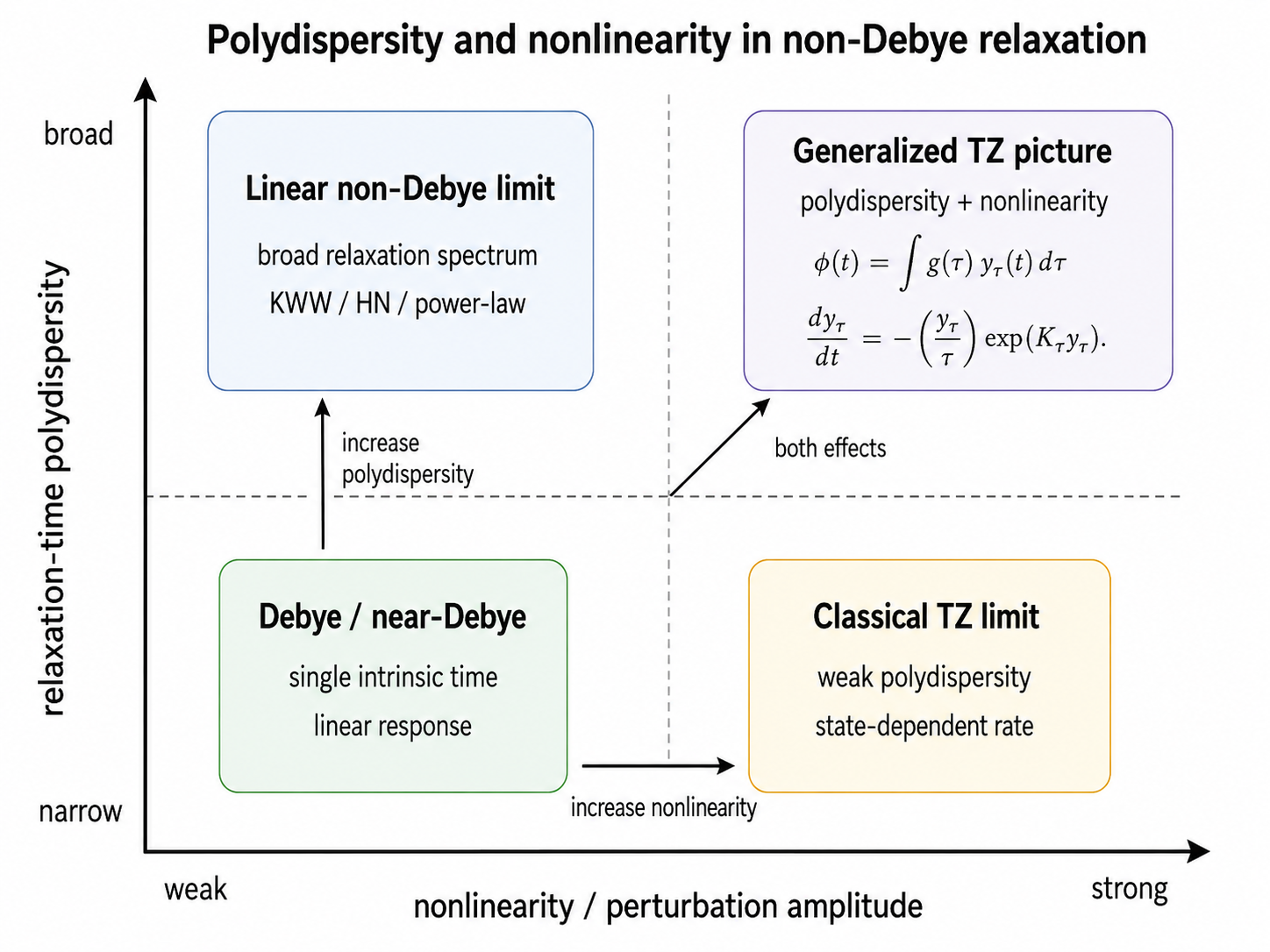}
\caption{Schematic classification of non-Debye relaxation in terms of two
distinct physical ingredients: relaxation-time polydispersity and nonlinear
state-dependent feedback. In the linear-response limit and for sufficiently
small perturbations, non-Debye relaxation may originate predominantly from a
distribution of relaxation times, giving KWW-like or power-law behaviour.
For systems with an approximately Debye-like intrinsic relaxation spectrum,
large perturbations may instead reveal predominantly nonlinear dynamics,
corresponding to the classical Trachenko--Zaccone mechanism. Realistic stress,
structural and dielectric relaxation may involve both effects simultaneously,
motivating a generalized TZ description combining a relaxation-time spectrum
with nonlinear feedback.}
\label{fig:polydispersity_nonlinearity}
\end{figure}

The classification illustrated in figure~\ref{fig:polydispersity_nonlinearity}
suggests a useful way of interpreting relaxation experiments. In the limit of a very small initial perturbation, isothermal stress or dielectric relaxation may remain within the linear-response regime. Non-Debye behaviour can then originate predominantly from polydispersity of relaxation times, giving rise, for example, to KWW-like or power-law relaxation without requiring appreciable nonlinear feedback. Conversely, in a system whose small-amplitude response is close to Debye relaxation, a sufficiently large initial perturbation may reveal nonlinearity even in the absence of substantial relaxation-time polydispersity. This corresponds most closely to the classical TZ picture, in which the instantaneous relaxation rate evolves with the state of the system.

Real relaxation processes may generally contain both effects. A natural extension of the TZ equation is therefore to associate a nonlinear TZ-type evolution with a distribution of relaxation modes. In a mean-field representation one may write
\begin{equation}
\phi(t)=\int_0^\infty g(\tau)\,y_\tau(t)\,d\tau,
\end{equation}
where $g(\tau)$ is a normalized distribution of relaxation times and each mode obeys
\begin{equation}
\frac{d y_\tau}{d t}
=
-\frac{y_\tau}{\tau}
\exp \left(K_\tau \, y_\tau\right).
\end{equation}
The classical TZ equation is recovered in the monodisperse limit
$g(\tau)=\delta(\tau-\tau_0)$, whereas the linear polydisperse limit follows for
$K_\tau\rightarrow0$. The general case contains both a spectrum of intrinsic
relaxation times and nonlinear state-dependent feedback.

This generalized viewpoint provides a possible bridge between two descriptions of non-Debye relaxation that are often treated separately. Relaxation-time polydispersity describes the coexistence of processes operating on different intrinsic time scales, whereas the TZ nonlinearity describes the modification of the rate of each process as relaxation proceeds. Their combination may therefore provide a more realistic description of stress, structural and dielectric relaxation in complex glass-forming materials.

An important experimental question is how these two contributions can be
disentangled. A particularly direct strategy would be to measure the same
relaxation process while systematically varying the amplitude of the initial
perturbation. In the linear-response limit, the relaxation spectrum should be
independent of perturbation amplitude, and deviations from Debye behaviour
would primarily reflect relaxation-time polydispersity. Nonlinear TZ feedback,
by contrast, should become increasingly apparent as the perturbation amplitude
is increased. Amplitude-dependent stress-relaxation, dielectric-relaxation or
structural-recovery measurements may therefore provide a direct experimental
test of the relative importance of polydispersity and nonlinear feedback.

This observation also suggests that the parameters $g(\tau)$ and $K_\tau$
should not necessarily be regarded as independent phenomenological quantities.
In a microscopic theory, both may ultimately originate from the same underlying
structural heterogeneity. Establishing such a connection would provide an
important route towards a microscopic generalized TZ theory.

The polydisperse nonlinear formulation above generalizes the TZ equation in
the space of relaxation modes. A complementary set of extensions arises when
one relaxes the assumptions of spatial homogeneity, constant parameters and
deterministic dynamics. We briefly outline these possibilities below.

\subsection{Spatially resolved and flux-coupled dynamics}
A first important direction concerns spatial heterogeneity. Introducing a
local field $y_i(t)$ for different regions and allowing exchange or migration
between them suggests a coupled form such as

\begin{equation}
\frac{d y_i}{d t}
=
-\frac{y_i}{\tau_i}
\exp(K_i y_i)
+
\sum_j M_{ij}y_j,
\label{eq:TZ_flux_discrete}
\end{equation}

where $M_{ij}$ represents fluxes, exchange, migration or coupling between
different regions. In a continuum description, the corresponding
generalization may take a reaction--diffusion form,

\begin{equation}
\frac{\partial y}{\partial t}
=
-\frac{y}{\tau}
\exp(Ky)
+
D\nabla^2 y,
\label{eq:TZ_reaction_diffusion}
\end{equation}

or, more generally,

\begin{equation}
\frac{\partial y}{\partial t}
+
\nabla\cdot{\bf J}
=
-\frac{y}{\tau}
\exp(Ky),
\label{eq:TZ_flux_continuum}
\end{equation}

with a constitutive relation for the flux ${\bf J}$. Such extensions would
allow the local nonlinear feedback characteristic of the TZ equation to
interact with transport, migration and spatial heterogeneity. They provide a
natural bridge to reaction--diffusion systems, propagating fronts, competing
populations and network dynamics.

\subsection{Time-dependent feedback and stochastic extensions}
A second direction concerns the parameters themselves. In the simplest
formulation, $K$ and $\tau$ are treated as constants over a given relaxation
experiment. In more complex systems they should instead be regarded as
effective quantities that evolve with temperature, pressure, aging,
environmental conditions or other slow variables. Allowing

\begin{equation}
K=K(t),
\qquad
\tau=\tau(t),
\label{eq:time_dependent_parameters}
\end{equation}

or introducing additional dynamical equations for these quantities would
transform the TZ equation into a coupled multiscale theory in which the
feedback mechanism itself evolves in time. Stochastic extensions are another
natural possibility, particularly for systems dominated by intermittency,
avalanches or finite-size fluctuations.

\section{Conclusions and outlook}
\label{sec:conclusions}

The Trachenko--Zaccone equation originated from a very specific problem in
condensed-matter physics: understanding how interactions between local
relaxation events can generate the strongly non-Debye dynamics observed in
supercooled liquids and glasses. Its central ingredient is remarkably simple.
The instantaneous rate is not fixed, as in Debye relaxation, but changes with
the state of the relaxing system itself. This nonlinear feedback is expressed
through

\begin{equation}
\frac{d y}{d t}
=
-\frac{y}{\tau}\exp(Ky),
\label{eq:conclusion_TZ}
\end{equation}

where the sign and magnitude of the dimensionless parameter $K$ determine how
the occurrence of previous events modifies subsequent dynamics.

From the microscopic viewpoint developed for local relaxation events, $K>0$
corresponds to feed-forward hindrance: earlier events increase the activation
barrier for later ones and progressively slow the relaxation. The resulting
dynamics is well represented by stretched-exponential relaxation. The
opposite sign, $K<0$, describes facilitating or avalanche-like dynamics in
which previous events lower the effective barrier for subsequent events,
giving rise to compressed-exponential relaxation. The non-interacting limit
$K=0$ recovers ordinary Debye relaxation. Stretched, Debye and compressed
relaxation therefore appear not as unrelated empirical laws but as different
regimes of the same nonlinear dynamical equation.

This feature distinguishes the TZ framework from descriptions in which a
particular functional form or exponent is specified from the outset. The
nonlinear parameter $K$ acts instead as a continuous dynamical coordinate
connecting slowing and accelerating regimes through the Debye point. The
mapping between $K$ and the effective KWW exponent $\beta$ provides a direct
connection between this nonlinear dynamical picture and the conventional
phenomenology of non-Debye relaxation.

An important development reviewed here is that the same mathematical
structure can emerge from physical arguments very different from the original
elastic feed-forward mechanism. In polymer glasses, the SL--TS2
thermodynamic framework leads independently to a TZ-type nonlinear evolution
equation for free volume, specific volume and enthalpy relaxation. This
derivation demonstrates that the exponential state-dependent rate need not
be regarded as a peculiarity of local elastic events in liquids. Rather, it
can arise whenever the evolving state of a system modifies the activation
scale controlling its subsequent dynamics.

The resulting perspective suggests that the most general content of the TZ
equation is a nonlinear rate-feedback principle. The microscopic
interpretation of $K$ is system dependent, but the mathematical mechanism is
the same: the state variable modifies its own instantaneous rate through an
exponential feedback. This observation provides a natural route for extending
the theory beyond spatially uniform mean-field relaxation.

The development of the TZ equation therefore suggests a hierarchy of
descriptions. At its origin lies the microscopic feed-forward mechanism of
interacting local relaxation events. Its mathematical formulation gives a
minimal nonlinear equation capable of interpolating continuously between
slowing and accelerating relaxation. Independent thermodynamic derivations
show that the equation is not tied uniquely to its original microscopic
interpretation, while spatial, stochastic and multicomponent extensions may
provide a route toward a substantially broader theory of nonlinear complex
systems.

In this sense, the lasting significance of the Trachenko--Zaccone equation
may lie less in any single application than in the simplicity of the
principle it expresses: dynamics can become non-Debye, non-exponential and
strongly nonlinear when the evolving state of a system continuously modifies
the rate of its own future evolution. The fact that this idea emerged from
discussions about local relaxation events in glasses and has subsequently
found manifestations in polymer thermodynamics and population dynamics is a
fitting reflection of Kostya Trachenko's approach to theoretical physics:
to search for simple physical mechanisms whose mathematical consequences
extend far beyond the problem in which they first appeared.

\section*{Acknowledgements}
This article is dedicated to the memory of Kostya Trachenko, whose physical intuition, clarity of thought, intellectual leadership and warm friendship were central to the development of the ideas reviewed here. We are grateful to colleagues and collaborators who have contributed to discussions on relaxation, glasses and nonlinear dynamics over the years, including Konrad Samwer, Alois Loidl, Peter Lunkenheimer, Birte Riechers, Beatrice Ruta, Simone Napolitano, Riccardo Casalini, Victor Yakovenko. \\

\bibliographystyle{iopart-num}
\bibliography{tzreview}

\end{document}